\documentclass[11pt]{article}
\usepackage[margin=1in]{geometry}
\usepackage{authblk}
\usepackage[dvipsnames]{xcolor}
\usepackage{hyperref}
\hypersetup{
colorlinks=true
,linkcolor=BrickRed
,citecolor=OliveGreen
,filecolor=Mulberry
,urlcolor=NavyBlue
,menucolor=BrickRed
,runcolor=Mulberry
,linkbordercolor=BrickRed
,citebordercolor=OliveGreen
,filebordercolor=Mulberry
,urlbordercolor=NavyBlue
,menubordercolor=BrickRed
,runbordercolor=Mulberry
,pdftitle={}
,pdfpagemode=FullScreen
}
\usepackage[utf8]{inputenc}
\usepackage{subcaption}
\usepackage{xargs}
\usepackage{graphicx}
\usepackage{xcolor}
\usepackage{amsmath}
\usepackage[version=4]{mhchem}
\usepackage{siunitx}
\usepackage{longtable,tabularx}
\usepackage{tabularray}
\UseTblrLibrary{booktabs,siunitx}
\usepackage{natbib}
\newcommand{\incfig}{\centering\includegraphics}
\newcommand\abs[1]{\left|#1\right|}
\newcommand{\alfadot}{\dot{\alpha}}

\usepackage[capitalise]{cleveref}

\title{A numerical demonstration of dynamic stall control}

\author{Sarasija Sudharsan and Anupam Sharma}
\affil{Iowa State University, Ames, IA, 50011}

\begin{document}
\maketitle

\section{Introduction}
\label{sec:introduction}
This work explores the use of dynamic stall onset criteria for stall control through numerical simulations.
Dynamic or unsteady stall is a critical phenomenon in aerodynamics for airfoils operating at high angles of attack, as it significantly impacts lift, drag, and overall performance.
It is characterized by the formation of a coherent vortex structure, namely, the dynamic stall vortex (DSV), which leads to large variations in the unsteady aerodynamic loads.
The DSV contributes to vortex-induced lift while it remains attached to the airfoil surface.
With increasing adverse pressure gradients, the DSV detaches and convects off the airfoil surface leading to a much more severe and persistent stall compared to the static case \citep{McCroskey1981}.
A crucial aspect of the dynamic stall control is initiating control action before the roll-up of the DSV since the flow field is more unstable and harder to control once the vortex has rolled up~\citep{Chandrasekhara2007}.
Mitigating stall is essential for improving aircraft efficiency and safety, as well as extending the operational envelope of various aerodynamic systems.
The two criteria used in this study are based on the Leading Edge Suction Parameter ($LESP$, \citep{Ramesh2014}) and the Boundary Enstrophy Flux ($BEF$, \citep{Sudharsan2022}).
These parameters have previously been shown~\citep{Ramesh2014,Narsipur2020,Sudharsan2022,Sudharsan2023} to be successful indicators of stall onset, since they exhibit critical behavior in the vicinity of stall.

Traditional stall mitigation methods rely on empirical models or fixed control laws~\citep{Sheng2005}.
In contrast, the present study provides a more direct, real-time, physics-based approach.
By adjusting the angle of attack in response to the onset of stall as indicated by the above stall criteria, a proactive means of controlling dynamic stall is presented.
The focus of this work is to provide a demonstration (proof of concept) of how these stall onset criteria can be used for stall mitigation.
The approach presented here can be augmented with more sophisticated machine learning techniques, for instance, reinforcement learning (RL)~\citep{Liu2025}, to achieve more advanced stall control.

The unsteady Reynolds-Averaged Navier-Stokes (uRANS) method is employed for numerical modeling of unsteady flow around a pitching airfoil undergoing dynamic stall, using the open-source code Stanford Unstructured (SU2)~\citep{SU2}.
The $LESP$ and $BEF$ parameters are tracked for an airfoil undergoing prescribed unsteady motion.
At each time step, a tracking parameter is calculated based on the current values of $LESP$/$BEF$ and a `reference' value determined from the prescribed motion.
When the tracking parameter reaches a threshold value, indicating the stall criterion has been met, the unsteady motion is smoothly adjusted to mitigate stall.


\section{Methods}
\label{sec:methods}

\subsection{Solver}
\label{sec:solver}
We solve the uRANS equations using the open-source code SU2~\citep{SU2}.
We have previously demonstrated~\citep{Sudharsan2023} that uRANS can capture the trends of $LESP/BEF$ variation and the relative instances of events such as lift stall and DSV formation with acceptable accuracy in the Reynolds number range considered.
SU2 solves the compressible Navier-Stokes equations in strong conservation form by discretization using a finite volume method, with an implicit, second-order, dual-time-stepping~\citep{jameson1991time} approach for time integration.
Convective fluxes are calculated using the low-dissipation Low Mach Roe ($L^2$ Roe)~\citep{osswald2016l2roe} model.
The two-equation SST $k-\omega$ turbulence model (version V2003m)~\citep{Menter2003}, with a freestream turbulence intensity level of 1\%, is used for closure.
The reference length and velocity scale for nondimensionalization are the airfoil chord, $c$, and the freestream velocity, $U_{\infty}$, respectively.
A normalized time step, $\Delta t = \Delta t_{\rm physical} \, U_\infty /c = 5 \times 10^{-4}$ is used for time integration.

\subsection{Grid \& Datasets}
\label{sec:grid_datasets}
\begin{figure}[htpb]
    \subcaptionbox{full view}{
        \incfig[width=0.32\linewidth]{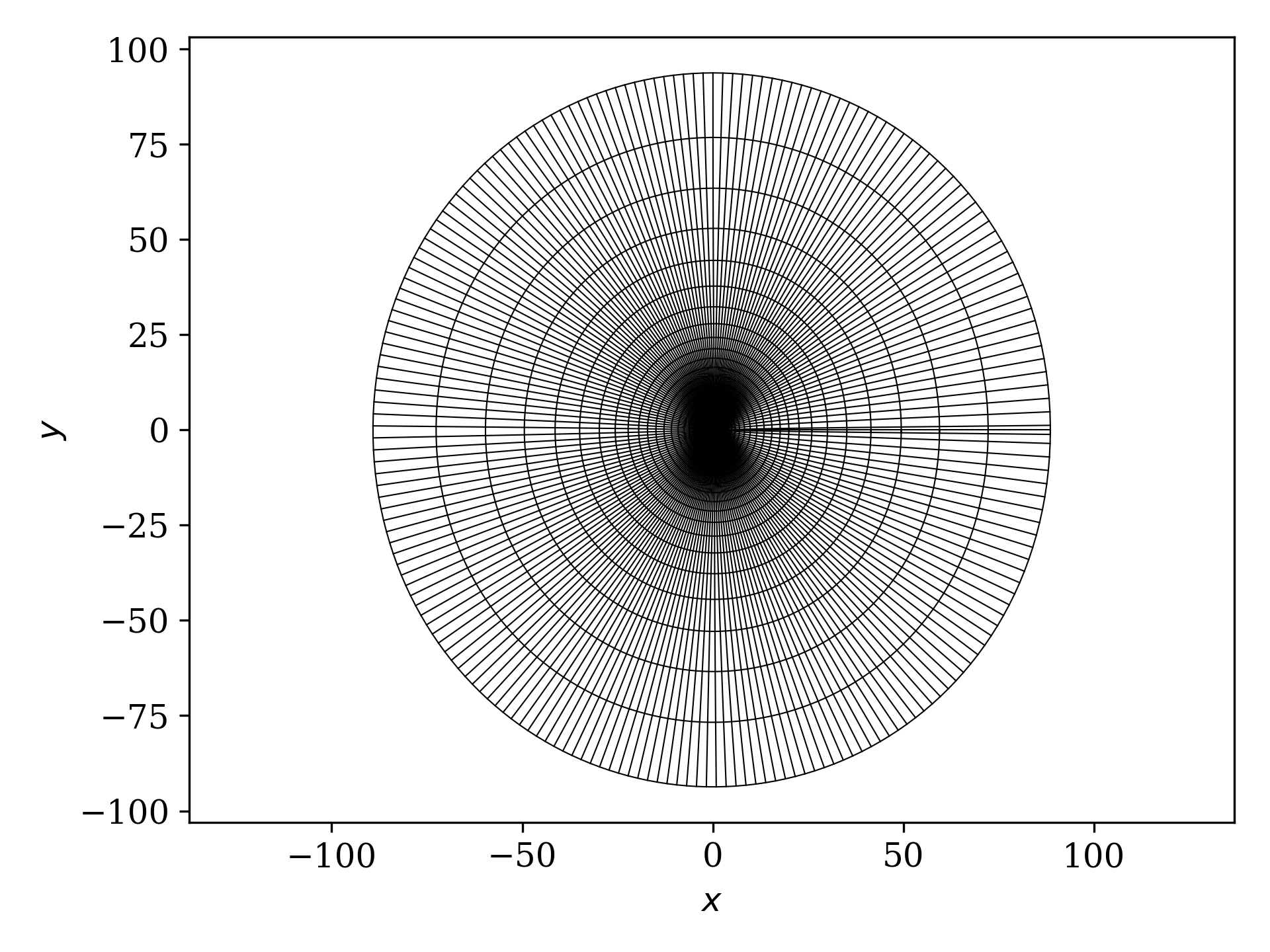}
    }
    \subcaptionbox{zoomed-in view}{
        \incfig[width=0.32\linewidth]{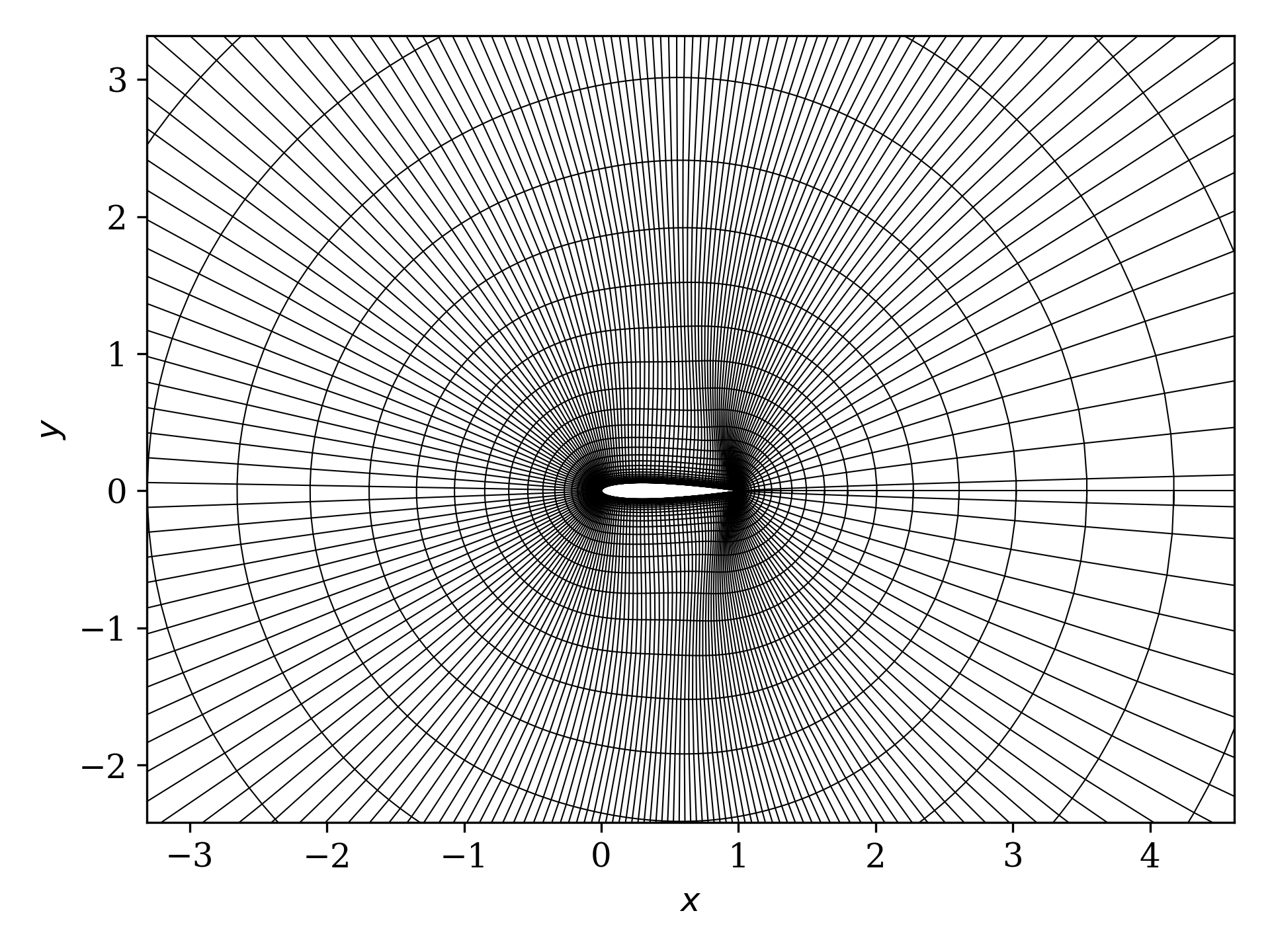}
    }
    \subcaptionbox{trailing-edge region}{
        \incfig[width=0.32\linewidth]{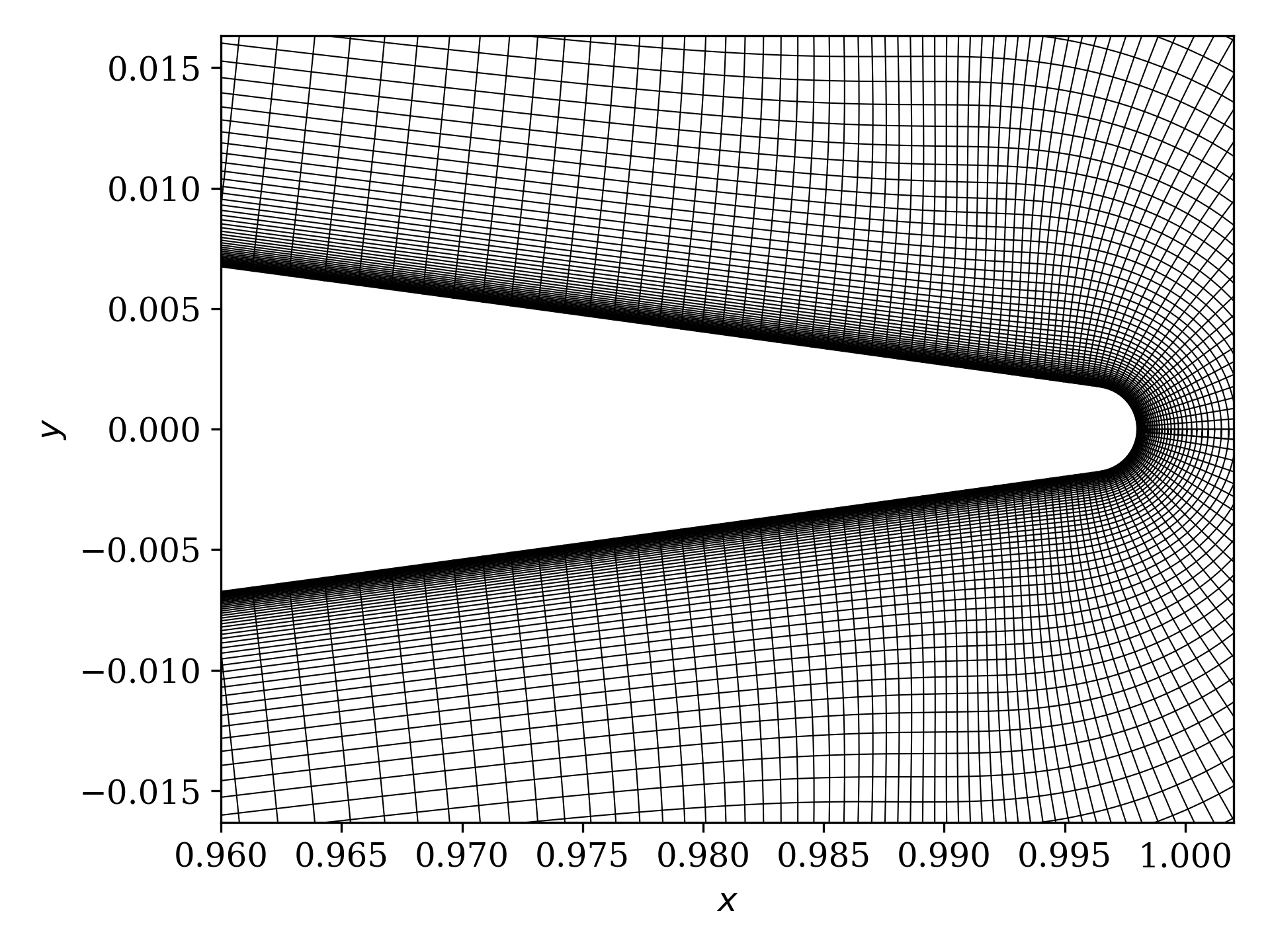}
    }
	\caption{Grid used in the present study: full view (left), zoomed-in view (middle) and trailing-edge region (right). Every third point in the radial and circumferential directions are shown for clarity in the left and middle panels.}
	\label{fig:grid}
\end{figure}

%
A structured O-Mesh having $598 \times 180$ grid points in the circumferential and radial directions, respectively, is used for all the simulations, with $y^+$ values less than $1$ over the entire airfoil surface.
\Cref{fig:grid} shows different views of the grid used for the simulations.
The solver setup and grids used herein remain similar to our prior work at these $Re$ ranges~\citep{Sudharsan2023}.
The grid was finalized based on comparisons with prior large-eddy simulations (LES)~\citep{Sharma2019,Sudharsan2022} and more details are available for reference in~\citet{Sudharsan2023}.

\begin{figure}[htbp]
    \incfig[width=0.7\linewidth]{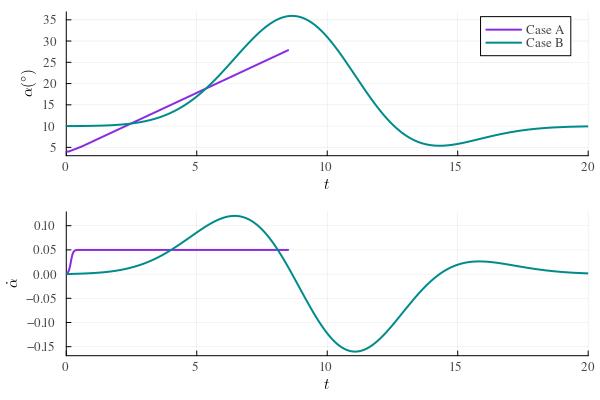}
    \caption{Variation with time of the angle of attack $\alpha$ for the two cases considered in the present work (top panel) and the corresponding pitch rate $\alfadot$ (bottom panel).}
    \label{fig:kinematics}
\end{figure}

The datasets used in the present work consist of uRANS simulations carried out at chord-based Reynolds numbers, $Re$, of $2 \times 10^5$, at a freestream Mach number, $M_{\infty}$, of 0.1 for an NACA 0012 airfoil.
The dynamic simulations are initiated from a static solution flow field at an initial angle of attack, $\alpha_0$. 
Two different unsteady motions (Cases A and B) are prescribed beginning from the static solution, as shown in \cref{tab:datasets} and \cref{fig:kinematics}.
The angular velocity of the unsteady motion is nondimensionalized by $U_{\infty}/c$ to yield a nondimensional pitch rate,  $\alfadot$.
\begin{table}[htbp]
    \centering
    \begin{tabular}{c|l}
    \textit{Case} & \textit{Airfoil motion / maneuver} \\
    \toprule
        A & Constant-rate, pitch-up\\
        B & Gaussian-modulated sinusoid\\
    \bottomrule
    \end{tabular}
    \caption{Airfoil kinematics considered in the present work}
    \label{tab:datasets}
\end{table}

Case A is a constant-rate, pitch-up case, where $\alfadot$ is smoothly ramped up to a value of $0.05$.
$\alpha$ increases from an initial value of  $\alpha_0 = 4^{\circ}$  up to about $27^{\circ}$.
Case B is a Gaussian-modulated sinusoid case, where $\alpha$ varies sinusoidally beginning from  $\alpha_0 = 10^{\circ}$  and reaching a maximum amplitude of $35^{\circ}$.
$\alfadot$ varies between about $\pm 0.15$.
\Cref{fig:kinematics} shows the variation with time, $t$, of $\alpha$ (top panel) and $\alfadot$ (bottom panel) for the two cases.
At $t = 0$, a hyperbolic tangent function is used to smoothly transition into the unsteady airfoil motion.

\subsection{Definitions of \texorpdfstring{$LESP$ and $BEF$}{LESP and BEF}}
\label{sec:lesp_bef_definitions}
The definitions of $LESP$ and $BEF$ are provided in this section for reference.
The focus of the present work is on demonstrating their use for stall mitigation.
As shown in \cref{eq:befdef}, $BEF$ is calculated by integrating the product of the wall vorticity and pressure gradient around the airfoil leading edge.
It is a strong measure of flow acceleration around the leading edge.
\begin{equation}
    BEF = \frac{1}{Re}\int_{x_p}^{x_s} \omega \frac{\partial \omega}{\partial n} \rm{d}s \approx - \int_{x_p}^{x_s} \omega \left(\frac{1}{\rho} \frac{\partial p}{\partial s} \right) \rm{d}s,
    \label{eq:befdef}
\end{equation}
where $\omega$, $p$, and  $\rho$ are the nondimensional wall vorticity, pressure, and density respectively. 
$n$ and $s$ are the normal and tangential coordinate directions to the airfoil surface.
The integral is carried out between some value of $x$ on the pressure side to that on the suction side; this value is set to $0.05$ based on prior work \cite{Sudharsan2022}.
All quantities are evaluated in the airfoil frame of reference.

$LESP$, given by \cref{eq:LESP_def}, is a measure of the chord-wise suction force $F_{\rm suction}$ near the leading edge, obtained by integrating surface pressure.
The integral to obtain $F_{\rm suction}$ is conventionally carried out from the maximum thickness point on the pressure side to that on the suction side~\citep{Ramesh2014}.
\begin{align}
    LESP = \sqrt{\abs{C_{\rm suction}}/(2 \pi)}, \;{\rm where} \; C_{\rm suction} &=  F_{\rm suction}/(q_{\infty}c), \; {\rm and} \; q_\infty = \rho_\infty U^2_\infty,\\
    {\rm and}
    \quad F_{\rm suction} &= q_{\infty} \int_{x_p}^{x_s} C_p \, \hat{e_n} \cdot \hat{e}_{\rm x}\, \rm{d}s.
\label{eq:LESP_def}
\end{align}

These two parameters demonstrate critical behavior when stall onset is imminent. 
The physical significance of the $LESP$ and $BEF$ parameters has been explored in prior work~\citep{Ramesh2014,Sudharsan2022}.
While these parameters can be readily evaluated using the full flow field data that is available through numerical simulations, we argue that these parameters can also be estimated using sparse data such as surface pressure taps and/or wall-shear measurements, as discussed in \citet{Sudharsan2022}.

\subsection{Tracking parameter and control strategy}
\label{sec:tracking_control}
A tracking parameter, $\sigma$, defined by \cref{eq:sigmadef} where $X$ is either $LESP$ or $\abs{BEF}$, is used to determine stall onset.
\begin{equation}
    \sigma_{X} = \left(\frac{X(t)}{X_{\rm ref}(t)} - 1.0\right)
    \label{eq:sigmadef}
\end{equation}
The reference value used in the definition of $\sigma$, with the subscript `ref', is set depending on the type of motion.
A threshold value of $\sigma$ ($=\sigma_{\rm ctrl}$) is used to determine when control is triggered.
$\sigma_{\rm ctrl}$ is set to $-0.05$, which represents the parameter dropping by 5\% from its reference value.
The $BEF$/$LESP$ parameters reaching their peak magnitude and dropping thereafter signifies imminent (stall) vortex formation, as illustrated in~\cref{sec:caseA}.
Therefore, the control action is triggered once $\sigma_{X}$ falls below this threshold value ($=-0.05$).

\begin{figure}[htb!]
    \incfig[width=0.8\linewidth]{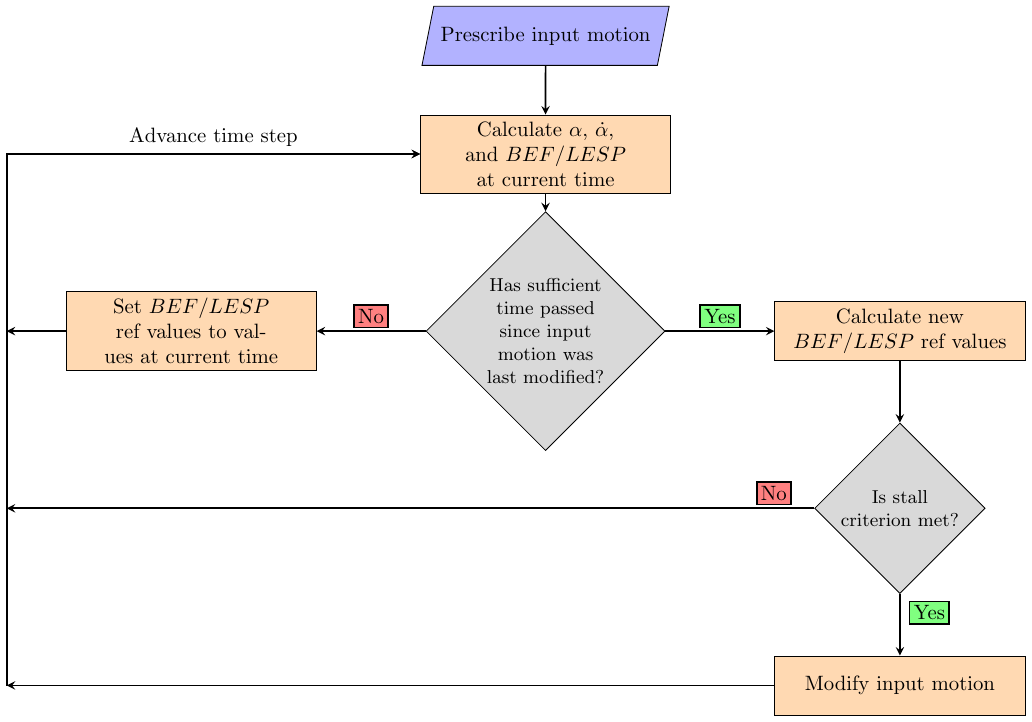}
    \caption{Flowchart showing the implementation of the control strategy for stall mitigation using $BEF$ and $LESP$.}
    \label{fig:flowchart}
\end{figure}

The flowchart in \cref{fig:flowchart} outlines the control strategy/approach.
At each time step, the angle of attack, the pitch rate, and the instantaneous values of $\abs{BEF}$ and $LESP$ are calculated.
Initially, the reference values for $\abs{BEF}/LESP$, (i.e., $\abs{BEF}_{\rm ref}$ or $LESP_{\rm ref}$) are set to small positive values.
After a few iterations to allow for transient effects to subside, the reference values are updated based on the time history of $\abs{BEF}$ and $LESP$.
The stall criterion, $\sigma(t) \le \sigma_{\rm ctrl}$, is assessed.
If the condition is met, the airfoil maneuver is modified to mitigate stall, otherwise, the simulation proceeds with the prescribed unsteady motion.

\section{Results \& Discussion}
\label{sec:results}

\subsection{Case A: Constant-rate, pitch-up motion}
\label{sec:caseA}
Case A involves the airfoil undergoing a constant-rate, pitch-up motion about the quarter-chord point. 
The rotation rate is smoothly ramped up using a hyperbolic tangent function to minimize numerical transients due to inertial (added mass) effects arising from acceleration ($\abs{\ddot{\alpha}} > 0$).

\begin{figure}[htpb]
    \centering
        \subcaptionbox{$\alpha = 10.0^{\circ}$}{\incfig[width=0.31\textwidth]{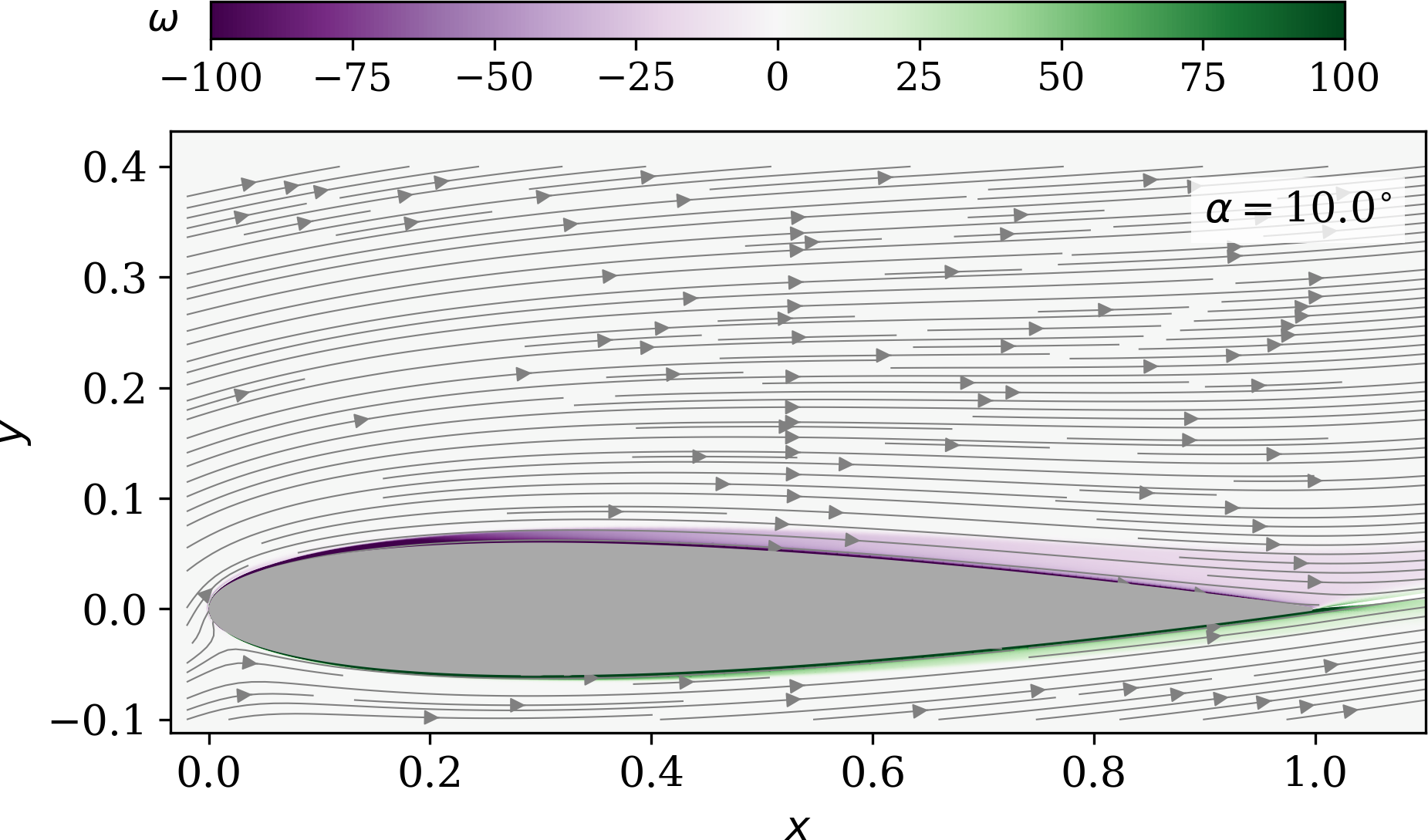}}
        \subcaptionbox{$\alpha = 12.0^{\circ}$}{\incfig[width=0.31\textwidth]{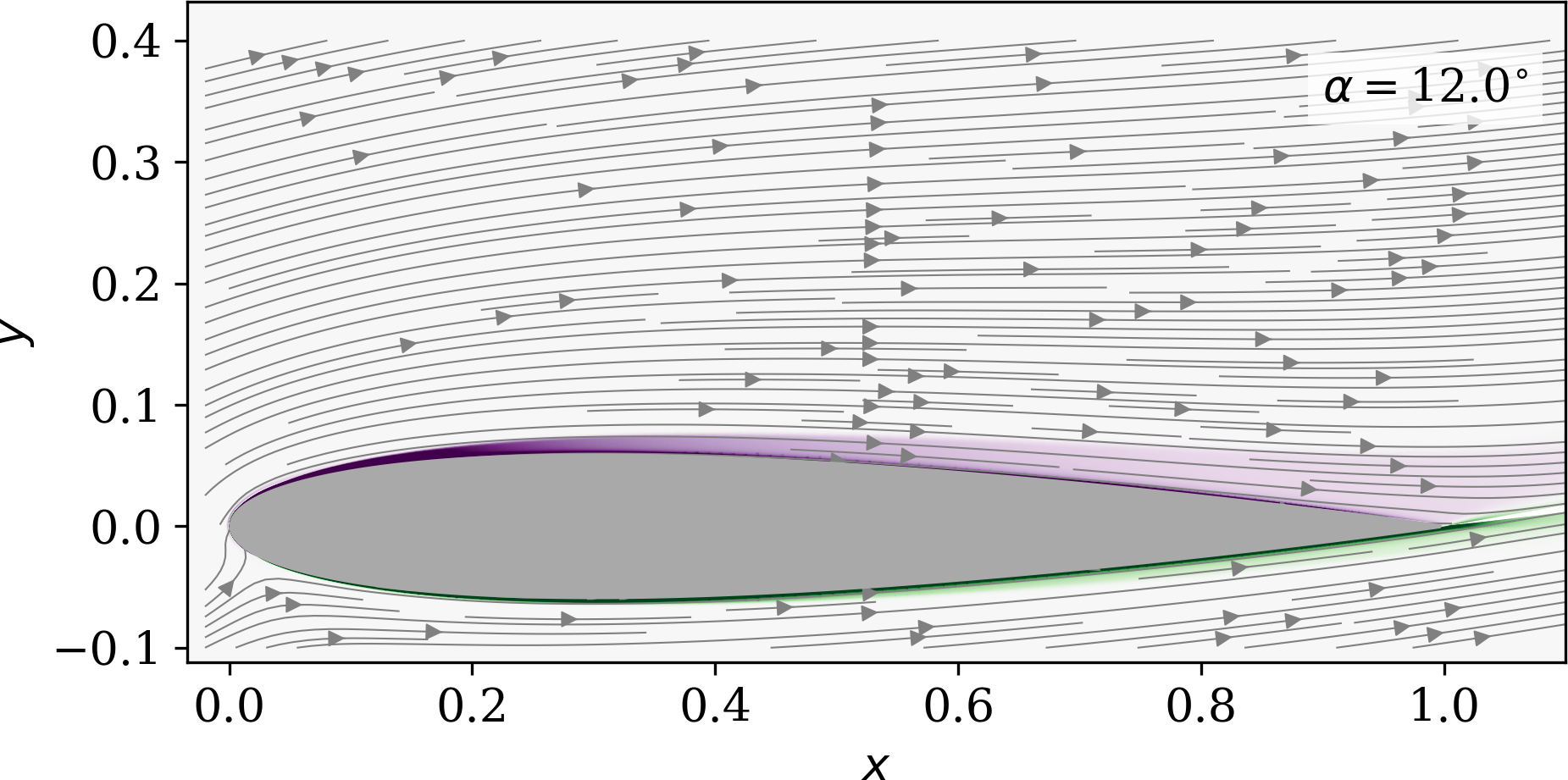}}
        \subcaptionbox{$\alpha = 15.0^{\circ}$}{\incfig[width=0.31\textwidth]{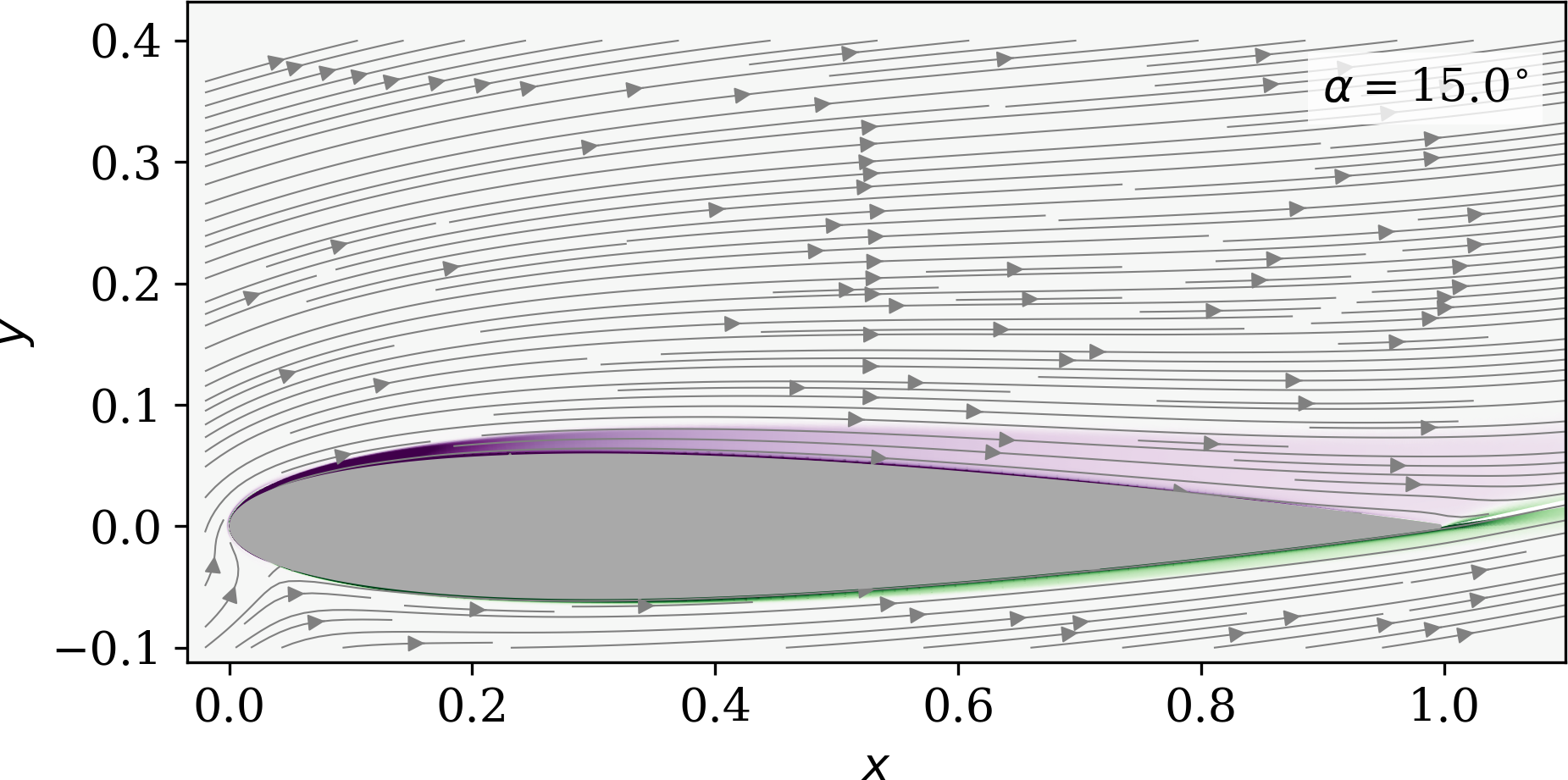}}
    \\
        \subcaptionbox{$\alpha = 16.8^{\circ}$}{\incfig[width=0.31\textwidth]{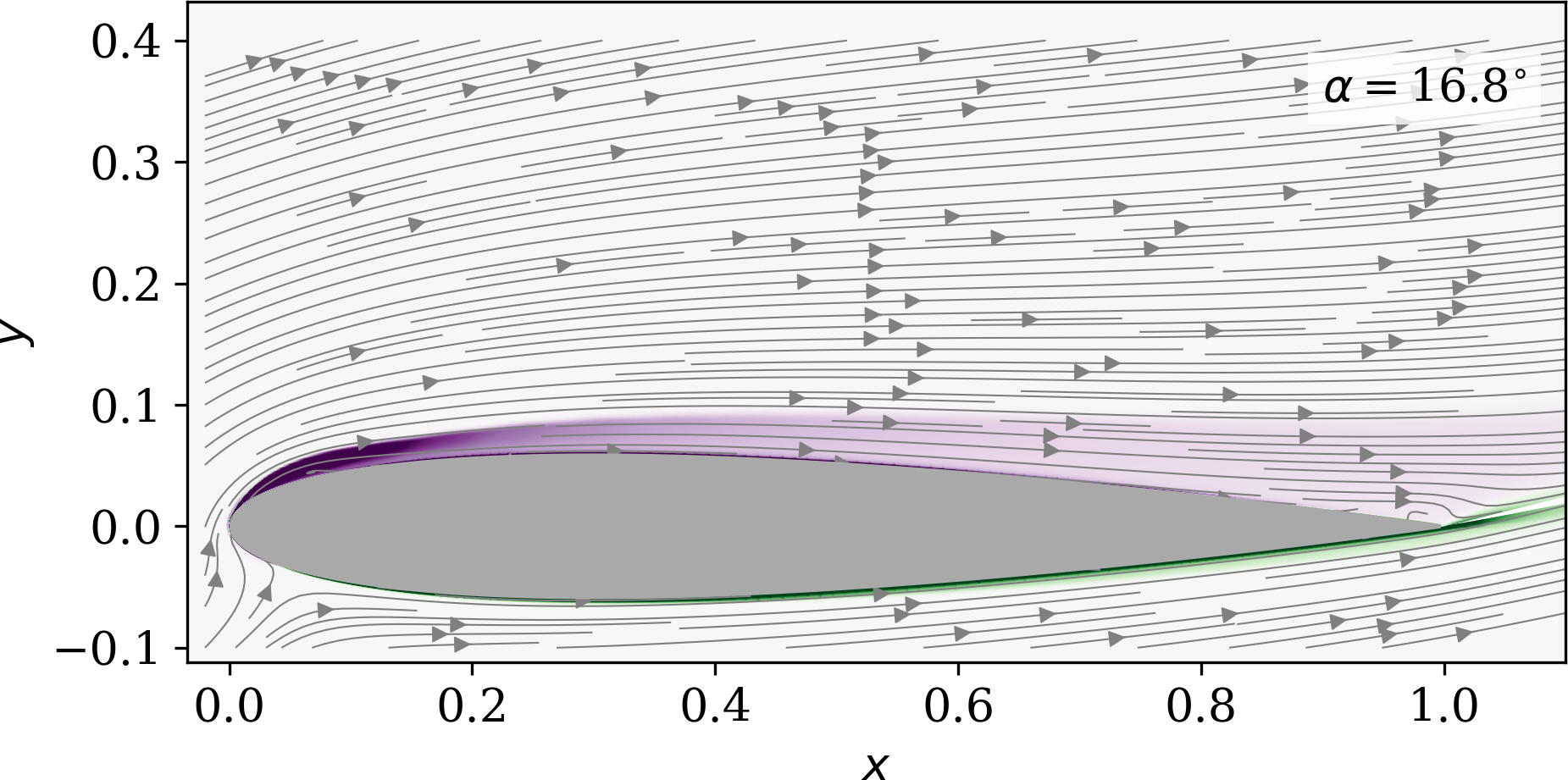}}
        \subcaptionbox{$\alpha = 17.5^{\circ}$}{\incfig[width=0.31\textwidth]{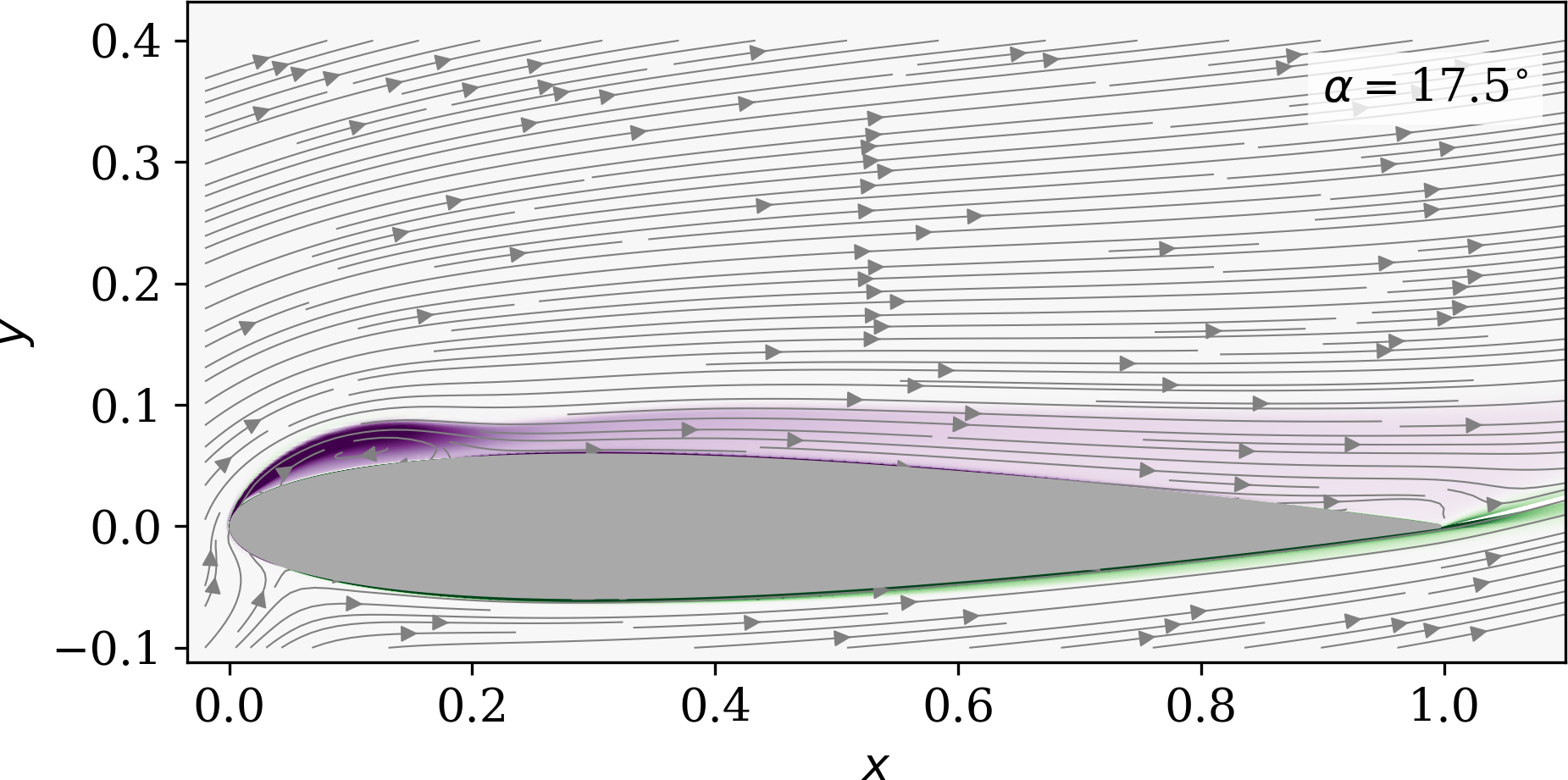}}
        \subcaptionbox{$\alpha = 18.9^{\circ}$}{\incfig[width=0.31\textwidth]{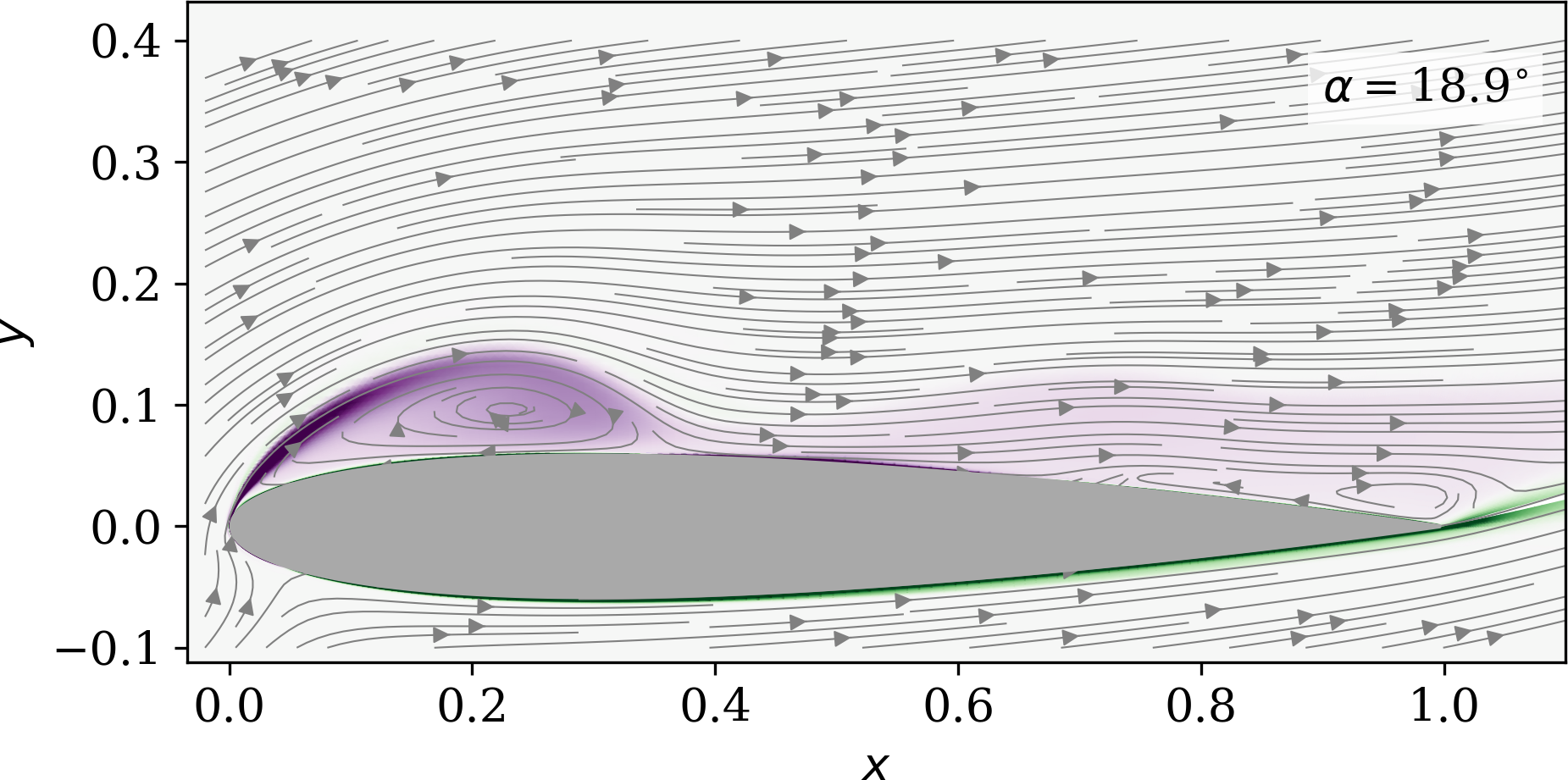}}
    \\
        \subcaptionbox{$\alpha = 21.0^{\circ}$}{\incfig[width=0.31\textwidth]{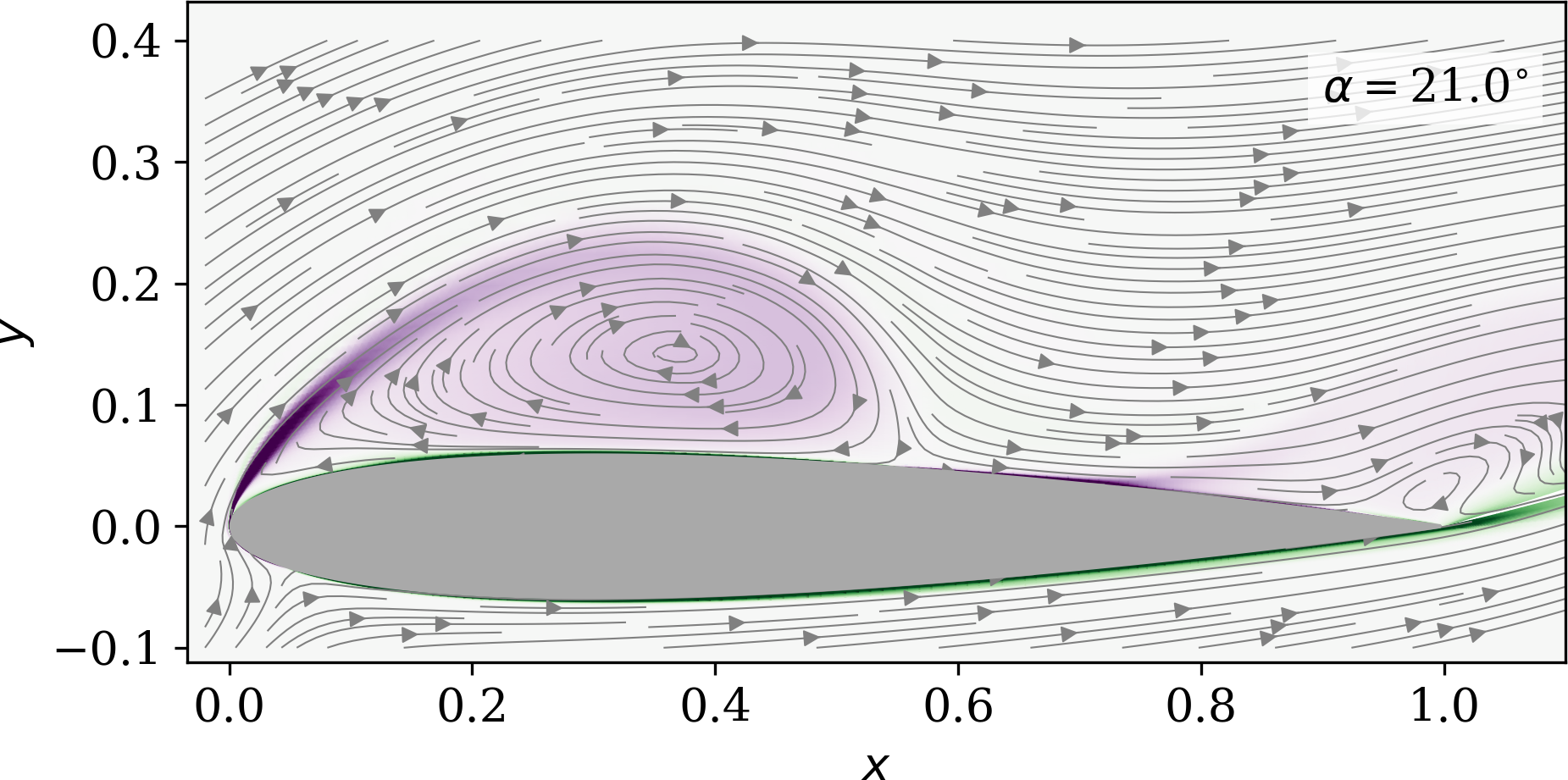}} \;
        \subcaptionbox{$\alpha = 22.5^{\circ}$}{\incfig[width=0.31\textwidth]{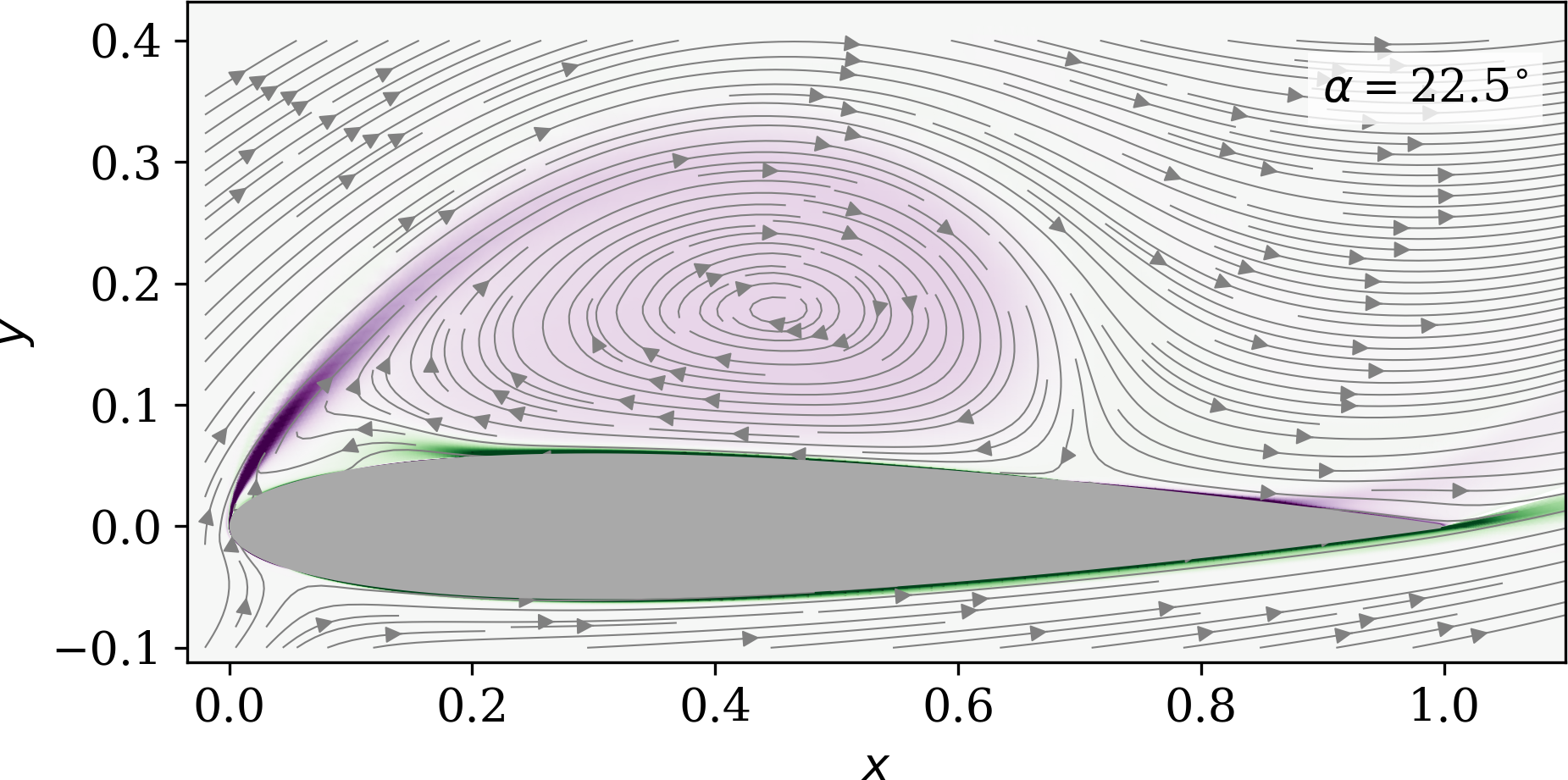}}
    \\
    \caption{Sequence of flow field snapshots for Case A when no control is applied (baseline); vorticity contours (legend in panel a) are shown overlaid with streamlines.
    \label{fig:flowfields_caseA}
    }
\end{figure}

\Cref{fig:flowfields_caseA} shows a sequence of snapshots of the flow field vorticity contours overlaid with streamlines in the airfoil frame of reference for the baseline (no control) case.
As the airfoil pitches up, the flow at the leading edge experiences an increasing adverse pressure gradient.
The accumulated vorticity near the leading edge coalesces into a coherent DSV
around $\alpha = 17.5^{\circ}$.
As the airfoil continues to pitch up, the DSV grows and convects downstream, while still remaining attached to the leading-edge shear layer.
While the vortex remains over the airfoil surface, it contributes to vortex-induced lift over the airfoil.
However, when the vortex convects off the airfoil surface (not shown), it leads to a large drop in the airfoil lift.

\begin{figure}[htpb]
    \incfig[width=0.7\textwidth,trim=0cm 0.8cm 0cm 1cm,clip=true]{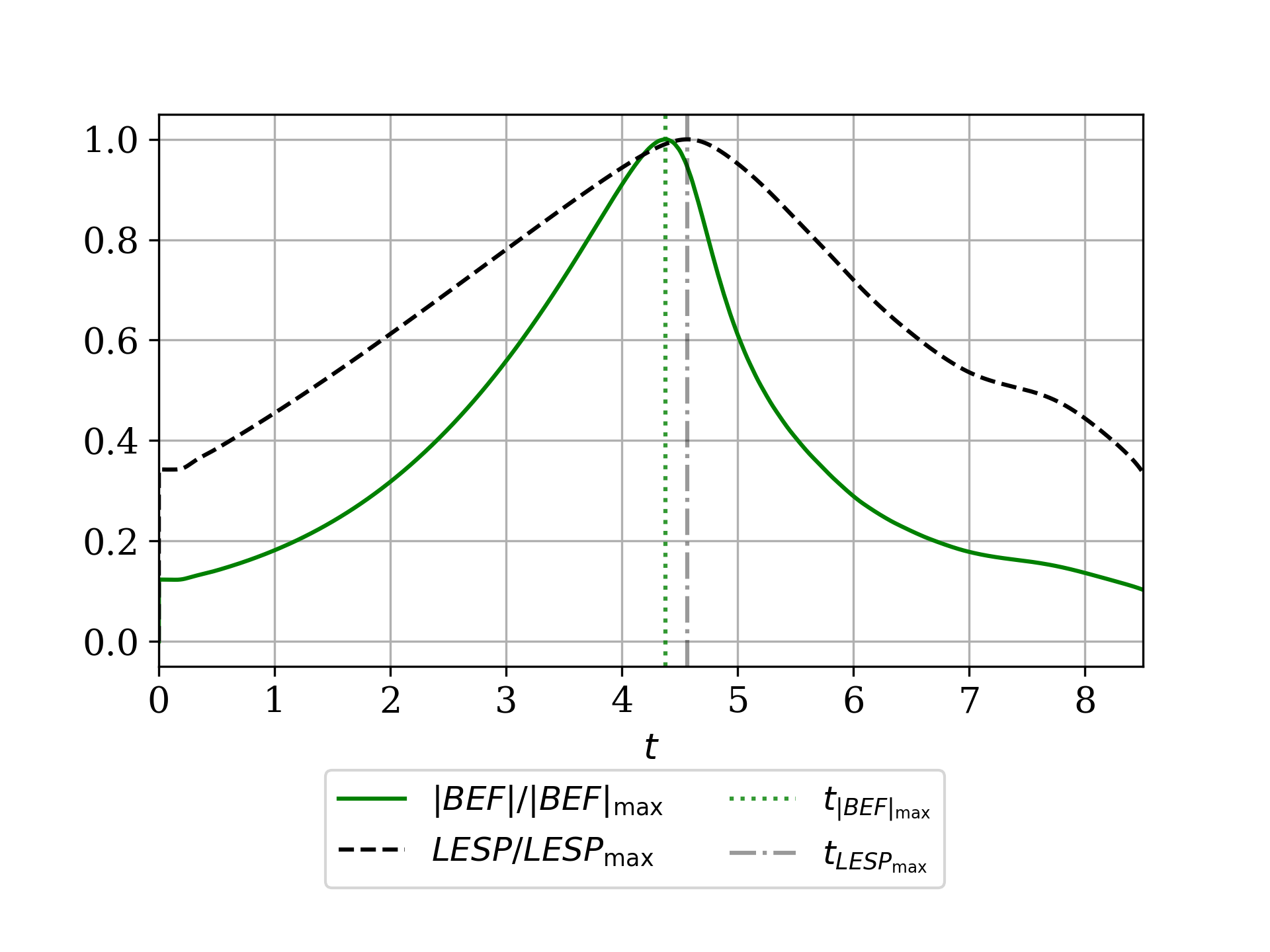}
    \caption{Variation with $t$ of $LESP$ and $\abs{BEF}$, each normalized by their respective maximum magnitudes reached during the unsteady motion, for Case A (baseline/no control).}
    \label{fig:lesp_bef_baseline_caseA}
\end{figure}

\Cref{fig:lesp_bef_baseline_caseA} shows the variation of $\abs{BEF}$ and $LESP$ with time; the parameters are normalized by their respective maximum values during the unsteady motion.
Both parameters demonstrate critical behavior in the vicinity of DSV formation, by reaching their maximum magnitudes.
For this case, the reference value in the definition of the tracking parameter, $\sigma$, is set to the maximum value of $\abs{BEF}$ and $LESP$ reached until that time.
That is, $\abs{BEF}_{\rm ref} (t) = \abs{BEF}_{\rm max} (t)$ in the definition (\cref{eq:sigmadef}) of  $\sigma_{\abs{BEF}}$, and $LESP_{\rm ref}(t) = LESP_{\max} (t)$ for $\sigma_{LESP}$. 

Once the stall criterion ($\sigma_X \le \sigma_{\rm ctrl}$) is met, $\alpha$ is smoothly reduced from the original, prescribed variation, $\alpha_{\text{prescribed}}(t)$, to a target value, $\alpha_{\text{target}}$;
$\alpha_{\text{target}}$ is set to $11^{\circ}$, which is slightly less than the static stall angle for the given conditions.
The start and end times of the control action are denoted by $t^*$ and $t^*_{\text{end}}$, respectively.
The function used to modify $\alpha$ after the criterion is triggered is given by \cref{eq:interpfn}.
\begin{equation}
    \alpha(t) = (1-w(\tau)) \cdot \alpha_{\text{prescribed}}(t) + w(\tau) \cdot \alpha_{\text{target}}, \quad t^* \le t \le t^*_{\text{end}},
    \label{eq:interpfn}
\end{equation}
where, $\tau = \left(t - t^*\right)/(t^*_{\text{end}} - t^*) \in [0,1]$, and $w(\tau) = 6\tau^5 - 15 \tau^4 + 10 \tau^3$ is a quintic smooth-step function, which is chosen such that $\dot{\alpha}$ and $\ddot{\alpha}$ are continuous when control is applied, and the control time duration, $t^*_{\text{end}}-t^* = 3$, is chosen to balance maximum acceleration and maximum $\alpha$ during the control maneuver.

\begin{figure}[htpb!]
    \incfig[width=0.75\textwidth,trim=0cm 0.8cm 0cm 1cm,clip=true]{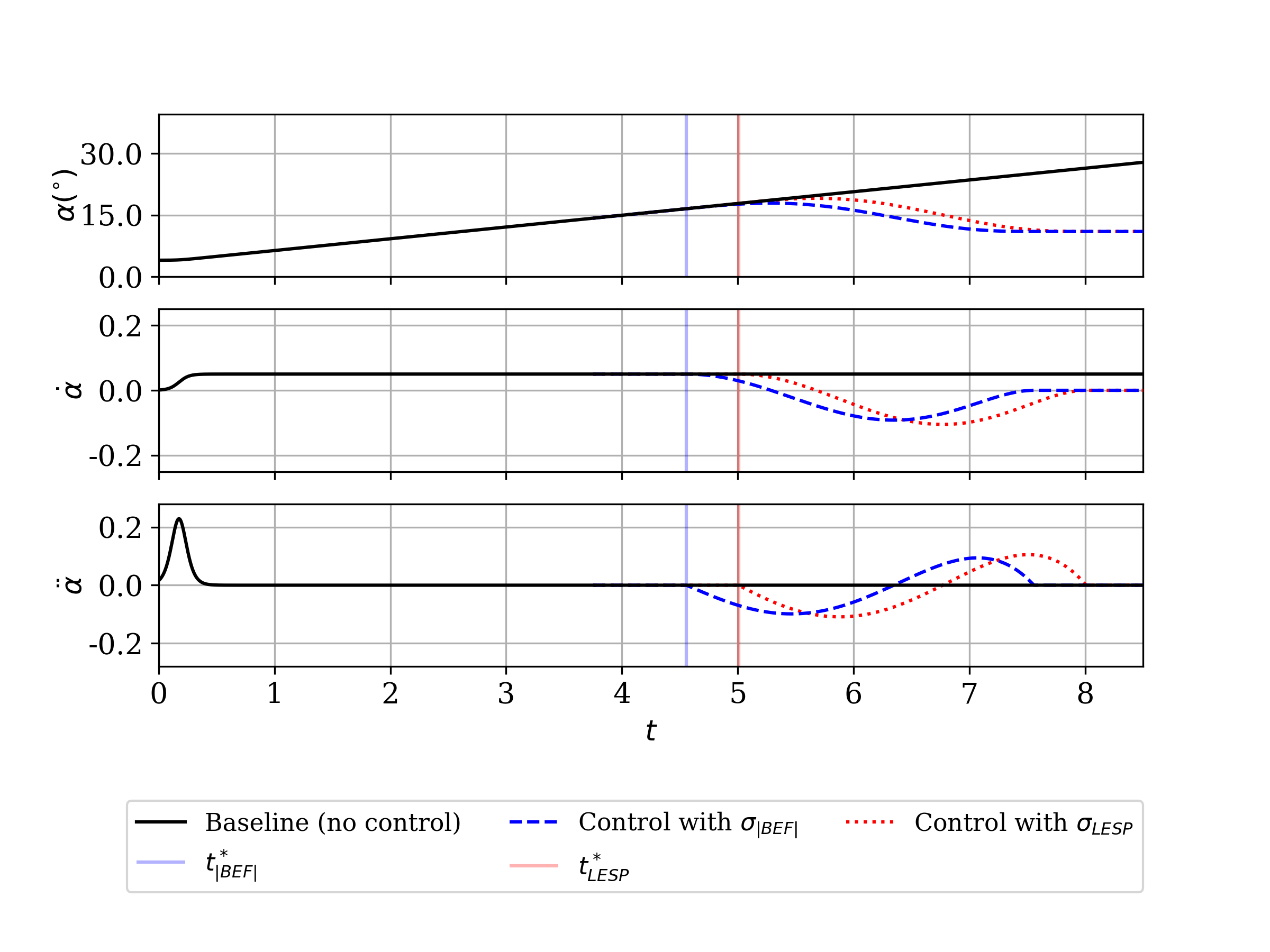}
    \caption{Variation with time of $\alpha$ (top panel) and $\alfadot$ (bottom panel) for Case A without any control applied (black), when control is applied using $\sigma_{\abs{BEF}}$ (blue), and using $\sigma_{LESP}$ (red).}
    \label{fig:alphavar_caseA}
\end{figure}

\Cref{eq:interpfn} ensures a smooth and continuous modification of $\alpha$ from $\alpha(t^*)$ to $\alpha_{\text{target}}$.
\Cref{fig:alphavar_caseA} shows the variation in $\alpha, \dot{\alpha}$, and $\ddot{\alpha}$ for the baseline case (no control applied), the case where  $\sigma_{\abs{BEF}}$ is used to trigger the control action, and the case when  $\sigma_{LESP}$ is used to trigger the control action.
The time instant when the control action is triggered while using $\abs{BEF}$ and $LESP$ are also indicated by vertical lines.
Since the $\abs{BEF}$ peak occurs earlier than the $LESP$ peak (see \cref{fig:lesp_bef_baseline_caseA}), the control action is triggered earlier when the $BEF$ parameter is used.

\begin{figure}[htpb!]
     \centering
     \hspace*{\fill}
        \subcaptionbox{Control using $BEF$, $t_{\abs{BEF}}^* \sim 4.5$\label{subfig:flowfield_bef_caseA}}{\incfig[width=0.49\textwidth,trim=0.5cm 0.2cm 0.7cm 0.2cm,clip=true]{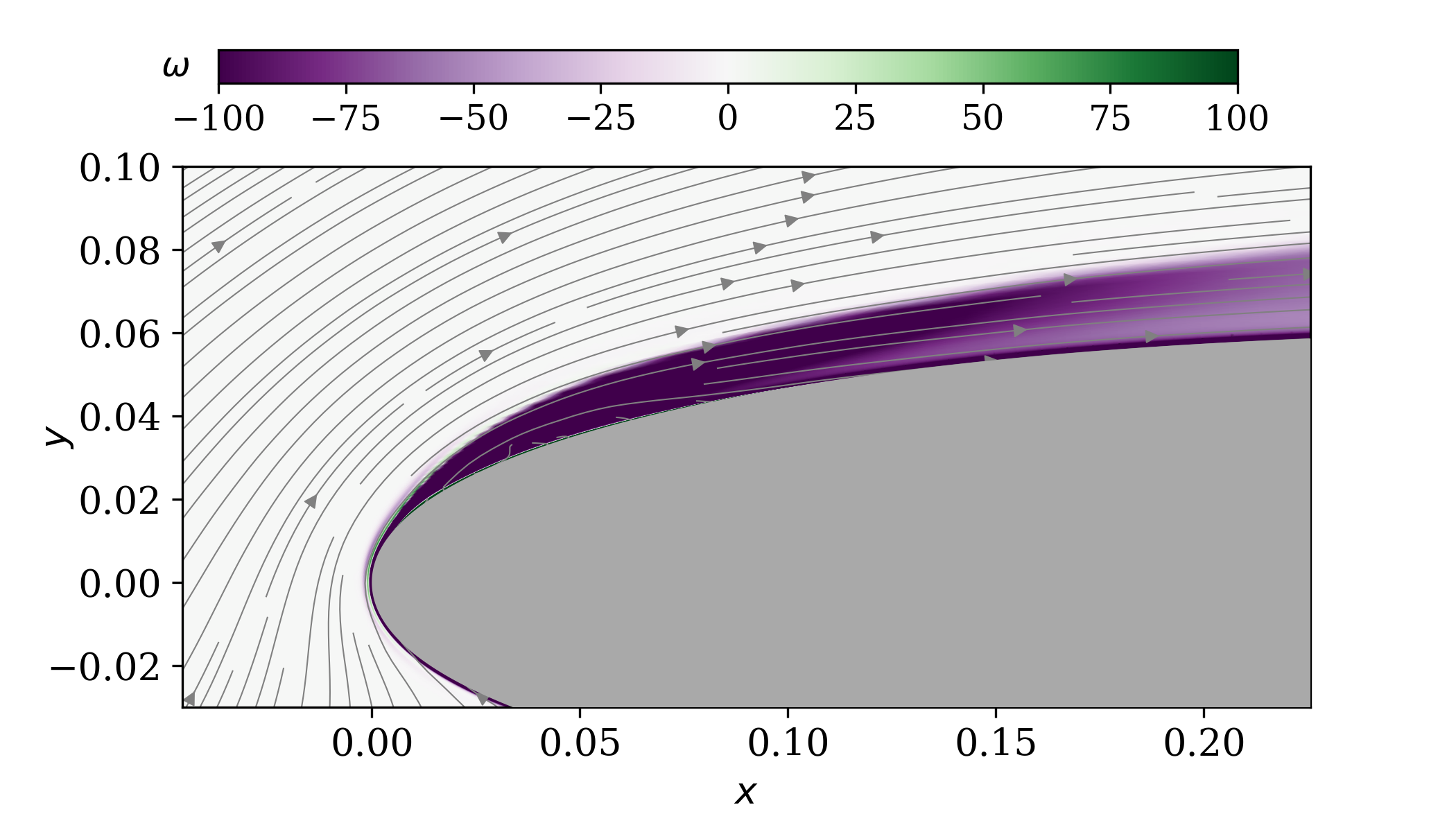}}
     \hfill
        \subcaptionbox{Control using $LESP$, $t_{LESP}^* \sim 5.0$\label{subfig:flowfield_lesp_caseA}}{\incfig[width=0.49\textwidth,trim=0.5cm 0.2cm 0.7cm 0.2cm,clip=true]{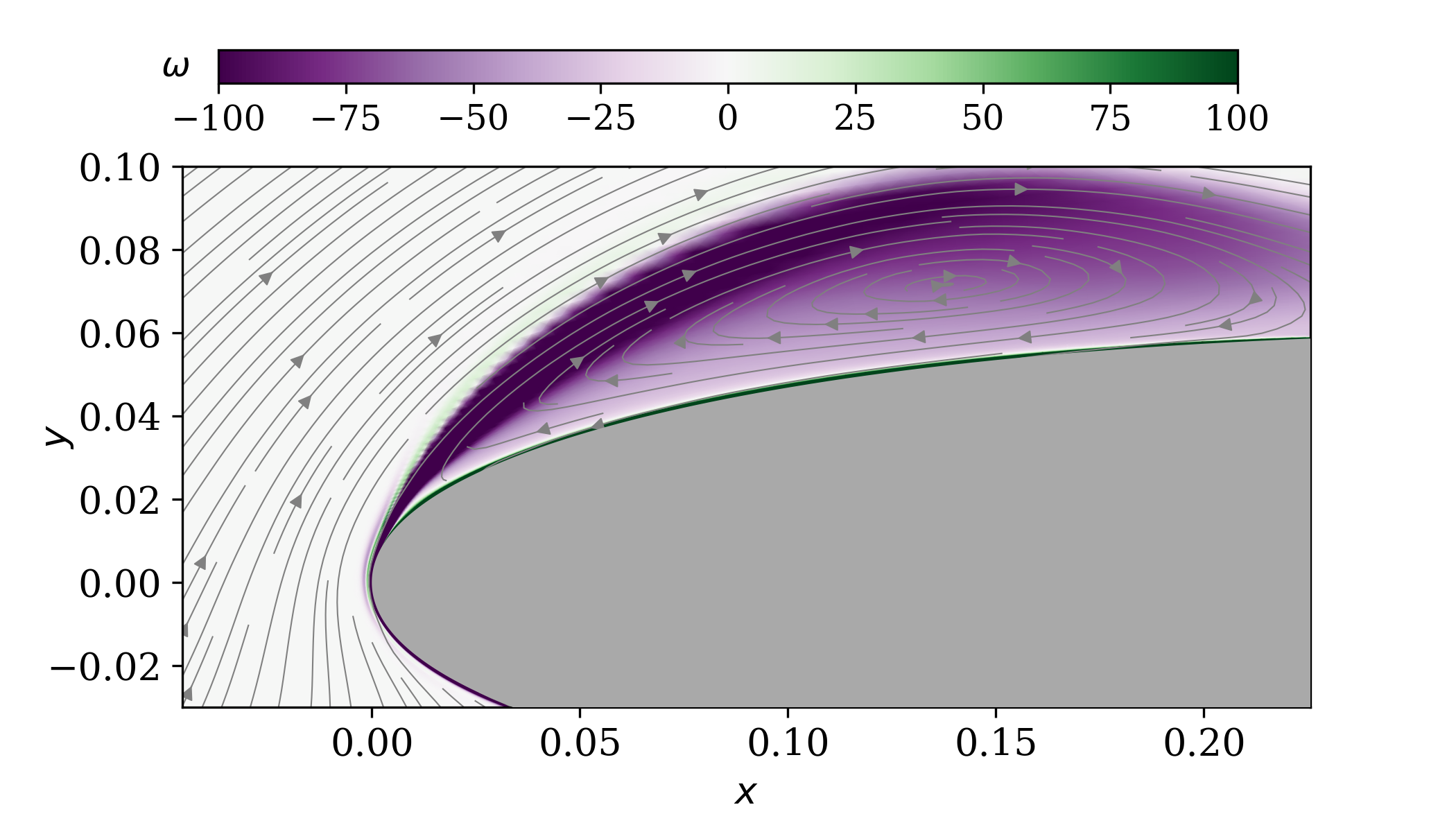}}
     \hspace*{\fill}
     \caption{Vorticity contours overlaid with streamlines near the airfoil leading edge at the instant when the stall criterion is triggered while using $\sigma_{\abs{BEF}}$ (a), and  $\sigma_{LESP}$ (b), respectively.}
    \label{fig:flowfields_ctrl_caseA}
\end{figure}

\Cref{subfig:flowfield_bef_caseA,subfig:flowfield_lesp_caseA} show the vorticity contours overlaid with streamlines at the time instants when control is applied using $\sigma_{\abs{BEF}}$ and $\sigma_{LESP}$, respectively.
While using the stall criterion based on $BEF$, the DSV is in a very early stage of development (\cref{subfig:flowfield_bef_caseA}).
In comparison, the DSV is much more developed while using the stall criterion based on $LESP$ (\cref{subfig:flowfield_lesp_caseA}).
Nevertheless, both criteria offer the opportunity to take stall mitigation actions before the complete development of the DSV, as apparent when comparing the flowfields in \cref{fig:flowfields_ctrl_caseA} with \cref{fig:flowfields_caseA}.

\begin{figure}[htpb!]
    \incfig[width=0.8\textwidth,trim=0cm 0.5cm 0cm 0.4cm,clip=true]{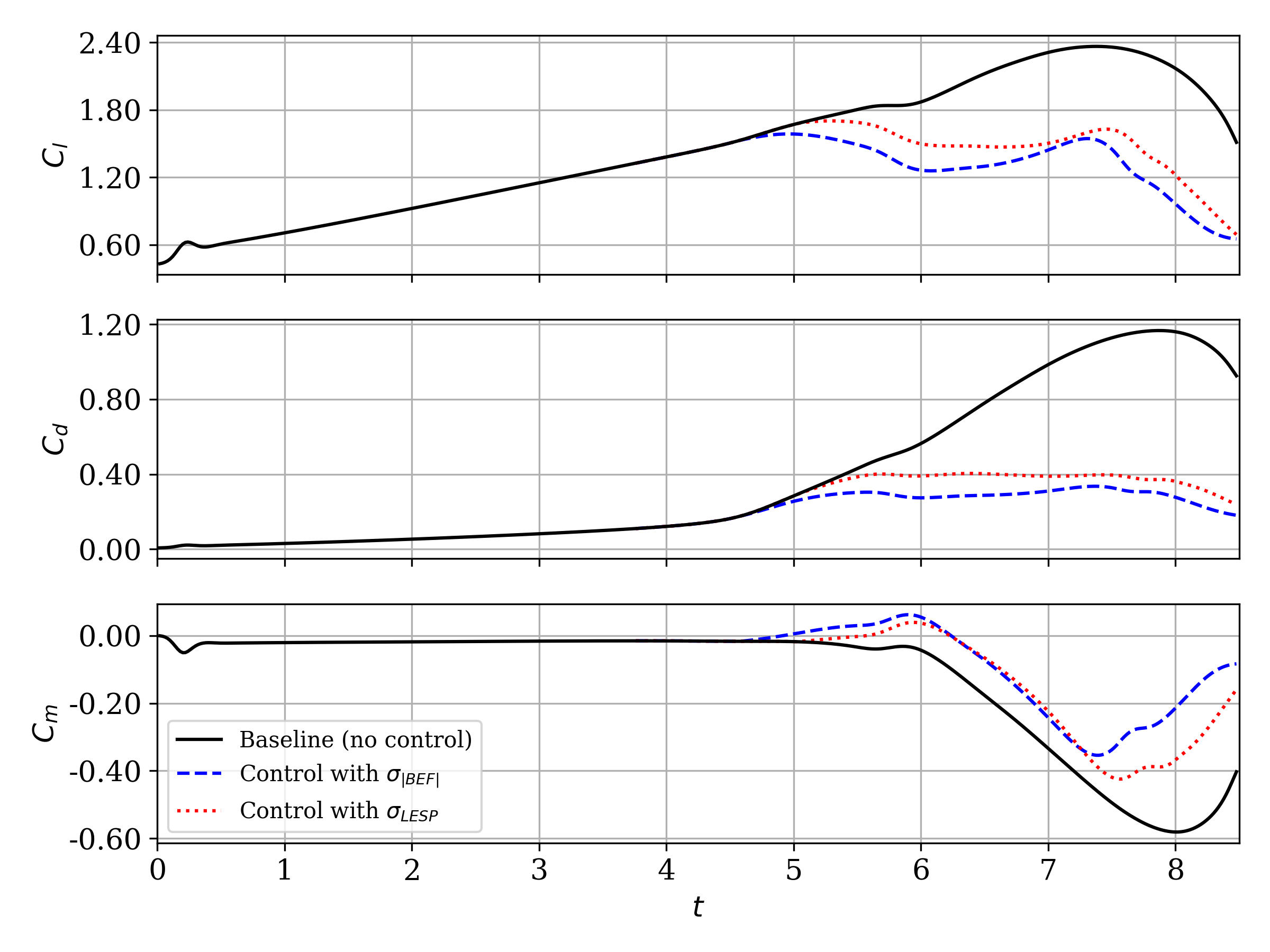}
    \caption{Variation with time of the unsteady aerodynamic coefficients ($C_l$,  $C_d$, and  $C_m$) for the baseline and control cases for Case A.}
    \label{fig:aerodyn_coeff_caseA}
\end{figure}

\Cref{fig:aerodyn_coeff_caseA} shows the time histories of the unsteady aerodynamic coefficients, ($C_l$, $C_d$, and $C_m$), for the baseline case and the two control cases using $\sigma_{\abs{BEF}}$ and $\sigma_{LESP}$.
Large unsteady loads associated with dynamic stall are averted in both control cases. 
The finite acceleration ($\ddot{\alpha}$) associated with changing the pitch rate results in a sharp but small jump in $C_m$ at the time of control application.
This is also associated with some vorticity shedding from the leading-edge, as indicated by the slight perturbation in $C_l$, however, these variations are small compared to those associated with eventual stall for the baseline case.
These effects can be controlled by adjusting the rate at which $\dot{a}$ is changed.
We emphasize that the intent of the present work is not to identify the best possible stall control strategy, but to demonstrate the feasibility of using $BEF$ and $LESP$ for stall mitigation.

\subsection{Case B: Gaussian-modulated sinusoid}
\label{sec:caseB}
The airfoil is pitched about the quarter-chord point and the pitch angle varied as a Gaussian-modulated sinusoid.
In the present case, the objective is to first mitigate stall and subsequently resume the original motion.
Unlike Case A where $\alpha$ was positive throughout the pitch-up maneuver, the stall criteria must now account for the potential onset of vortex formation and stall for both positive and negative values of $\alpha$.
The reference values used to evaluate the tracking parameter, $\sigma_X$ (\cref{eq:sigmadef}), are chosen based on whether the pitch rate contributes to an increase in the magnitude of $\alpha$ (see \cref{eq:refvalues_caseB}).
Accordingly, in addition to the condition that $\sigma_X$ falls below 5\%, an additional requirement, $\alpha \dot{\alpha} > 0$, is imposed.
This ensures that 
(a) the stall criterion works during both the pitch-up and pitch-down phases of the motion, and 
(b) the control action is triggered only when the unsteady pitch rate acts to increase the magnitude of $\alpha$.
\begin{equation}
    X_{\rm ref}(t) =
    \begin{cases}
        X_{\max}(t) & \text{if~} \alpha \, \dot{\alpha} > 0\\
        X(t) & \text{otherwise}.
    \end{cases}
    \label{eq:refvalues_caseB}
\end{equation}

Similar to Case A, a quintic smooth-step function, $w(\tau)$, is used to smoothly transition $\alpha(t^*)$ to $\alpha_{\text{target}}$ once the stall criteria are met.
When $\alpha$ reaches $\alpha_{\text{target}}$ at $t=t^*_{\text{end}}$, $\alpha$ is modified using a logistic blending function  $s(\gamma)$ so that the original motion is gradually resumed.
These functions are given in \cref{eq:ctrlfunc_caseB}.
\begin{equation}
    \alpha(t) =
    \begin{cases}
        (1-w(\tau)) \cdot \alpha_{\text{prescribed}}(t) + w(\tau) \cdot \alpha_{\text{target}}, & t^* \le  t <  t^*_{\text{end}}\\
        (1- s(\gamma)) \cdot \alpha(t^*_{\text{end}}) + s(\gamma) \cdot \alpha_{\text{prescribed}}(t), & t \ge  t^*_{\text{end}}
    \end{cases}  
    \label{eq:ctrlfunc_caseB}
\end{equation}

\noindent where, $s(\gamma) = 1/\left(1+ \exp(-\gamma)\right)$ is a logistic blending function. 
$\gamma = (t - (t^{*}_{\text{end}}))/L + k$, where $L=0.4$ controls the steepness of the transition, and  $k = \log \left( (1- \epsilon)/\epsilon \right) $ is a parameter that matches the end point values to within a small value, $\epsilon=1E$-$6$.

\begin{figure}[htpb]
    \incfig[width=0.75\textwidth,trim=0cm 0.8cm 0cm 1cm,clip=true]{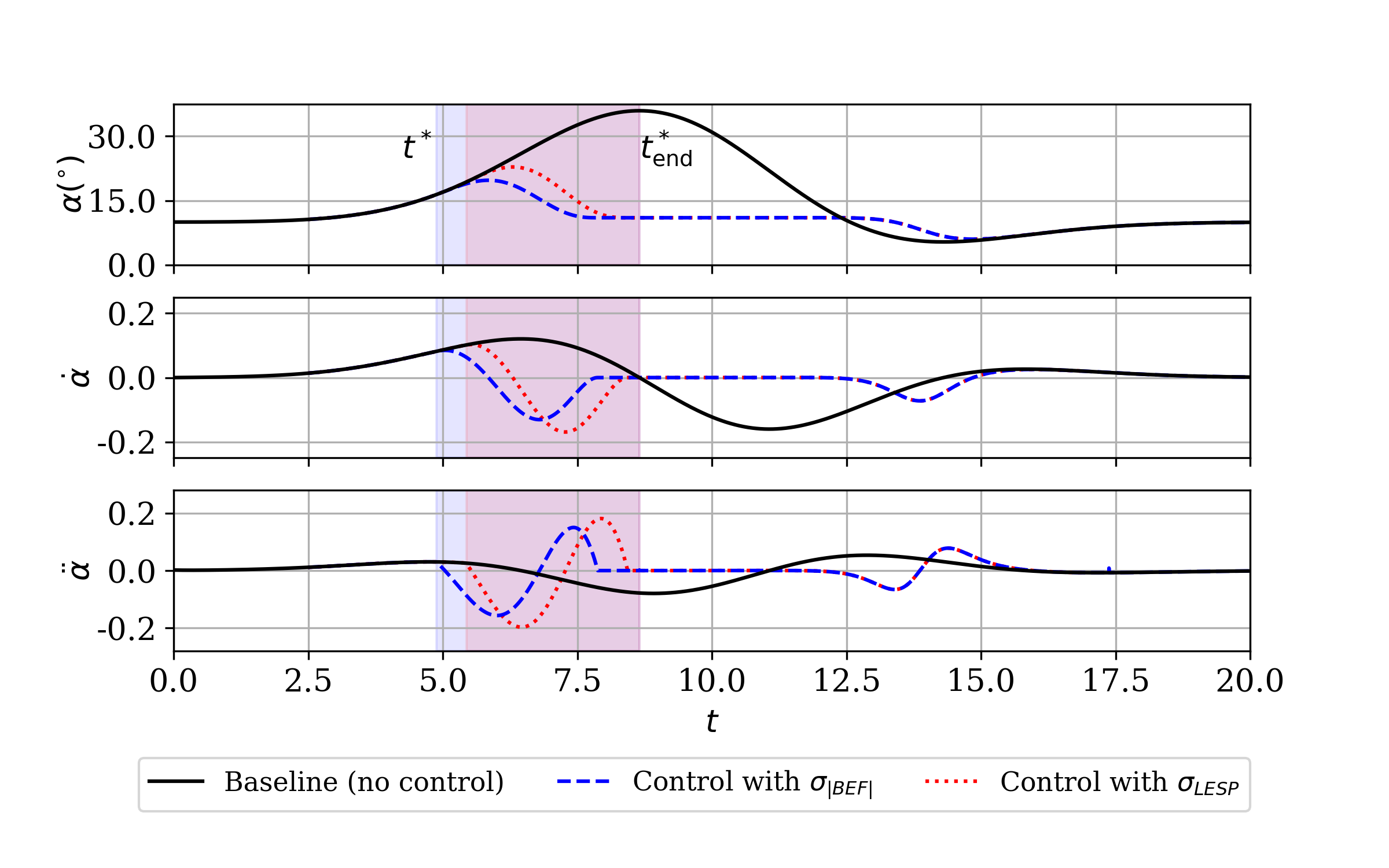}
    \caption{Time histories of $\alpha,~\dot{\alpha}$ and $\ddot{\alpha}$ for Case B with no control (baseline, in black), and with control using $\sigma_{\abs{BEF}}$ (blue) and $\sigma_{LESP}$ (red).}
    \label{fig:alphavar_caseB}
\end{figure}

\Cref{fig:alphavar_caseB} shows the time histories of $\alpha$ and $\dot{\alpha}$ for the baseline case without any control, and control implemented using $\sigma_{\abs{BEF}}$ and $\sigma_{LESP}$.
Once the criterion is triggered around $t^*$ ($\sim 5$ for $\sigma_{\abs{BEF}}$), $\alpha$ is smoothly reduced to $\alpha_{\text{target}}$.
Subsequently, $\alpha$ is changed to gradually resume the original motion, beginning at $t^*_{\text{end}} \sim 8.5$.

\begin{figure}[htpb]
    \centering
    \subcaptionbox{Baseline, $t=6$}{\incfig[width=0.31\textwidth,trim=0.8cm 0.9cm 0.9cm 0.6cm,clip=true]{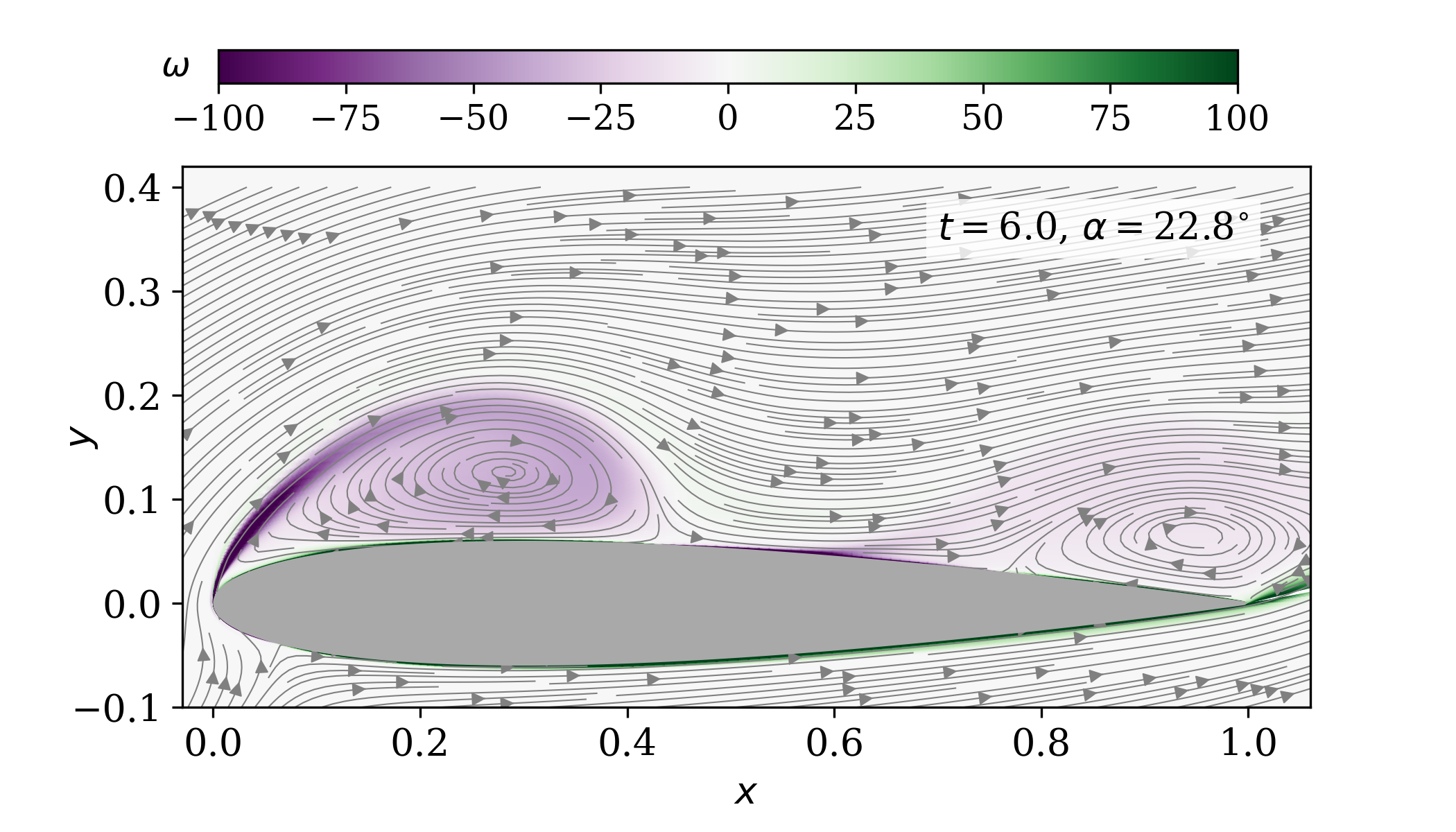}}
    \subcaptionbox{Control using $BEF$, $t=6$}{\incfig[width=0.31\textwidth,trim=0.8cm 0.9cm 0.9cm 0.6cm,clip=true]{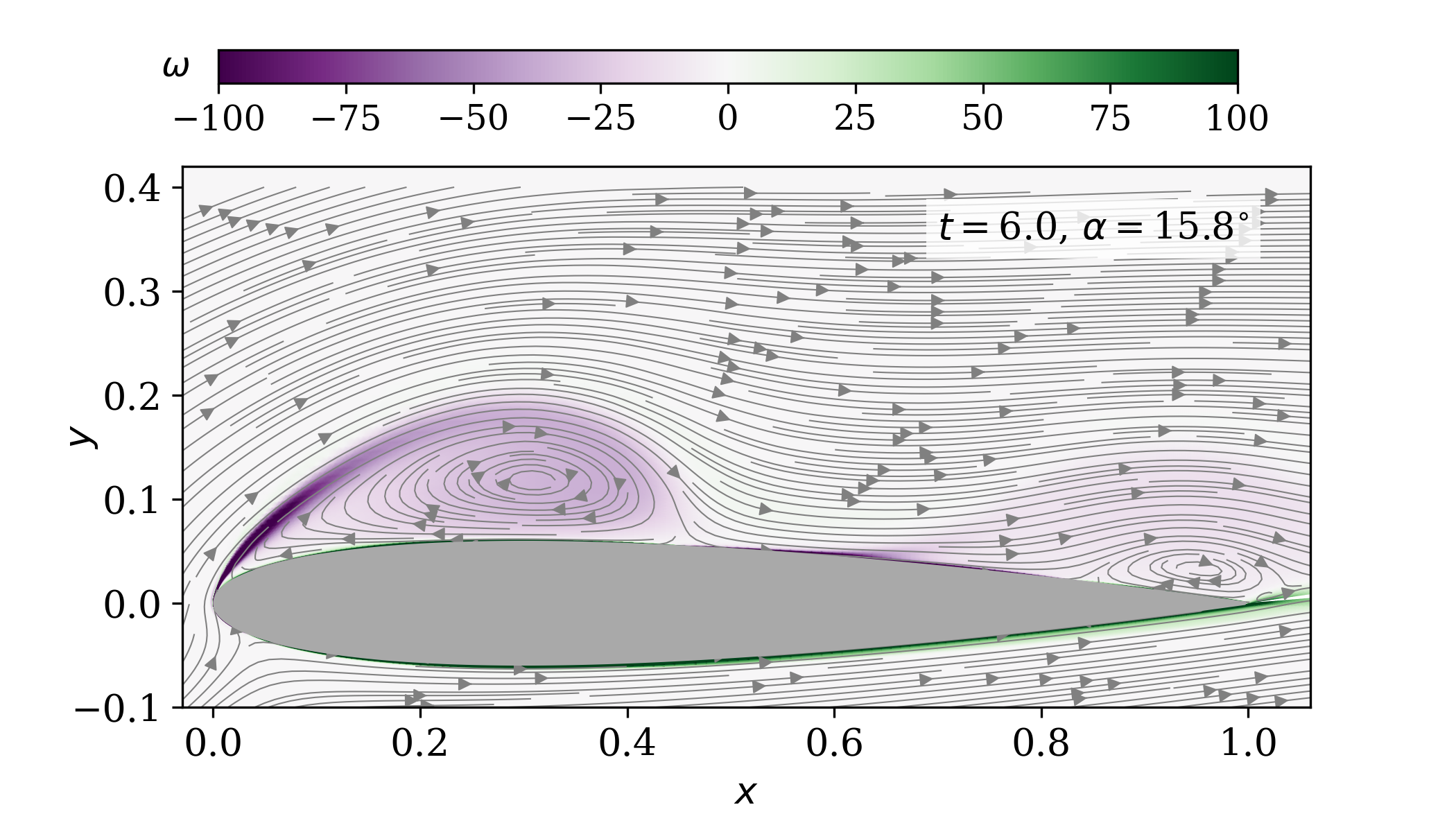}}
    \subcaptionbox{Control using $LESP$, $t=6$}{\incfig[width=0.31\textwidth,trim=0.8cm 0.9cm 0.9cm 0.6cm,clip=true]{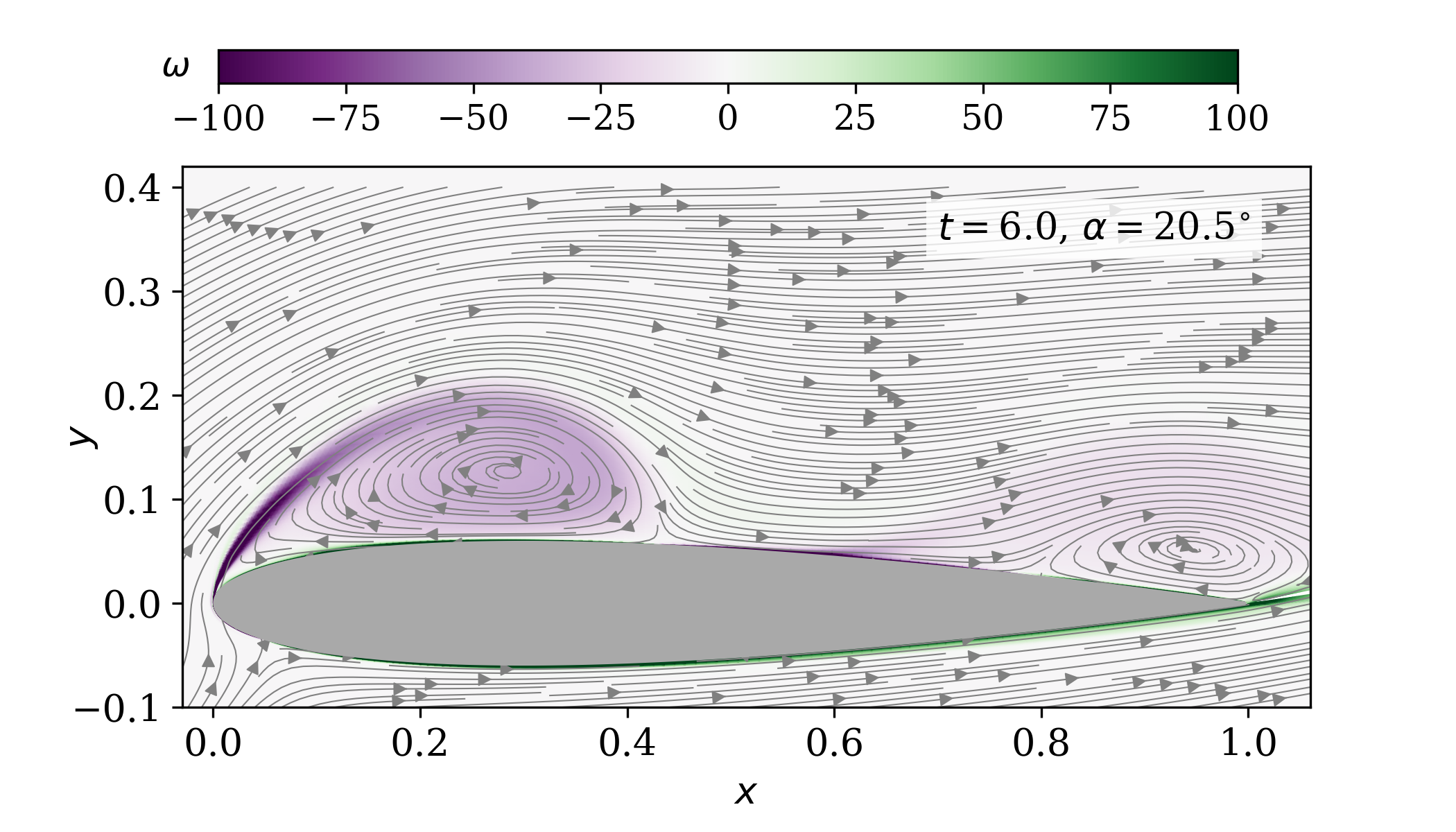}}
    \\
    \subcaptionbox{Baseline, $t=7.8$}{\incfig[width=0.31\textwidth,trim=0.8cm 0.9cm 0.9cm 0.0cm,clip=true]{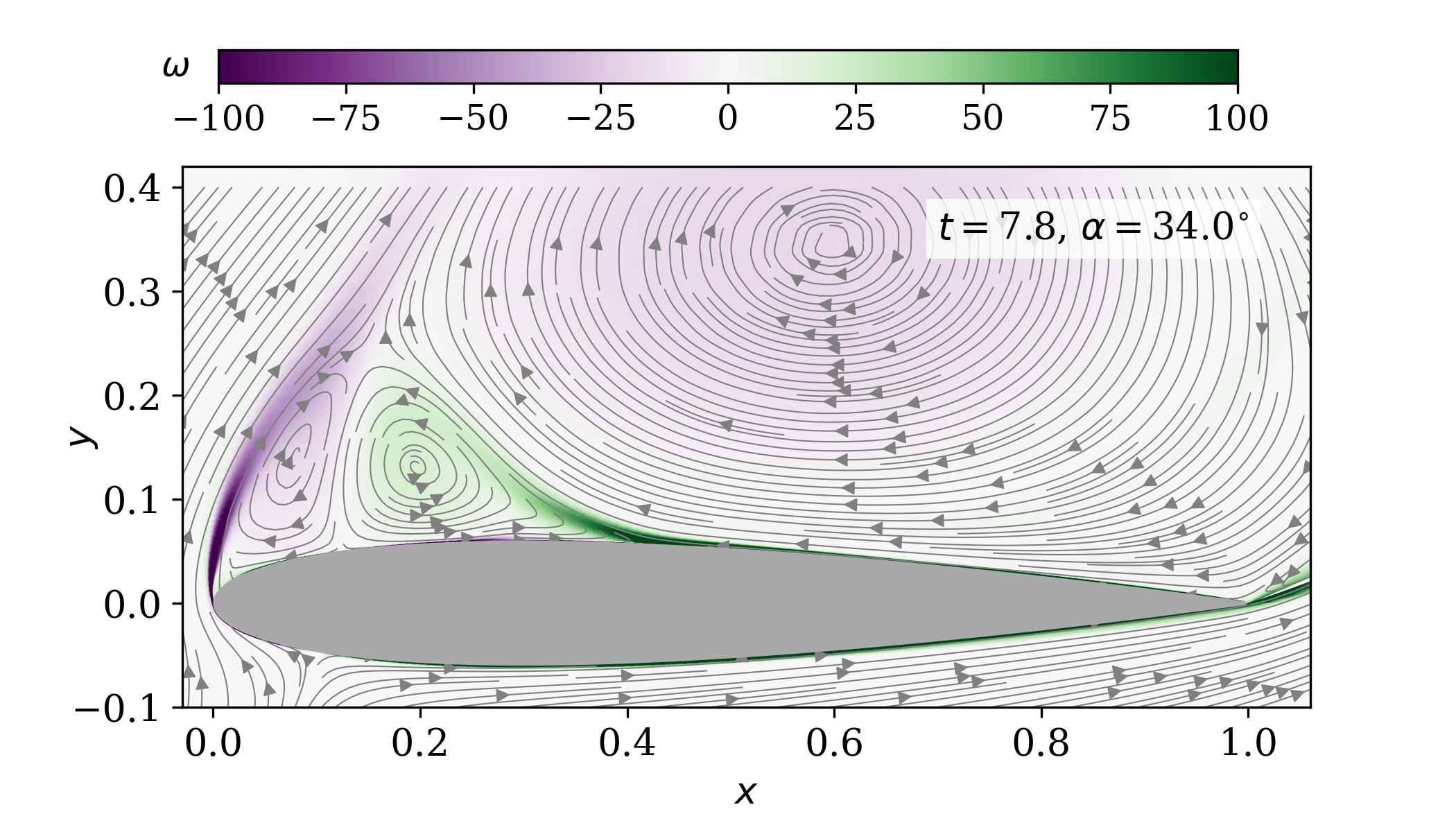}}
    \subcaptionbox{Control using $BEF$, $t=7.8$}{\incfig[width=0.31\textwidth,trim=0.8cm 0.9cm 0.9cm 0.0cm,clip=true]{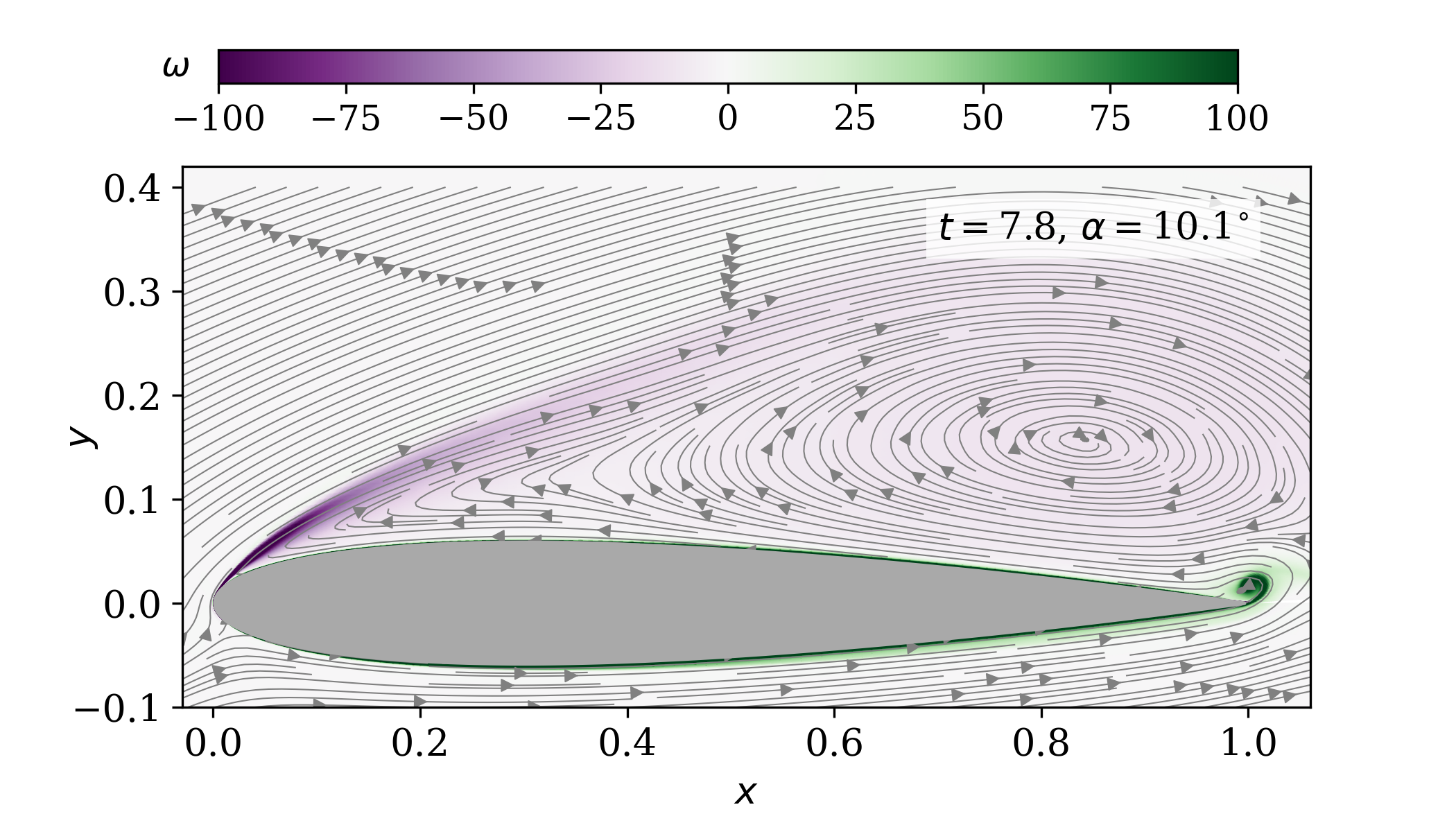}}
    \subcaptionbox{Control using $LESP$, $t=7.8$}{\incfig[width=0.31\textwidth,trim=0.8cm 0.9cm 0.9cm 0.0cm,clip=true]{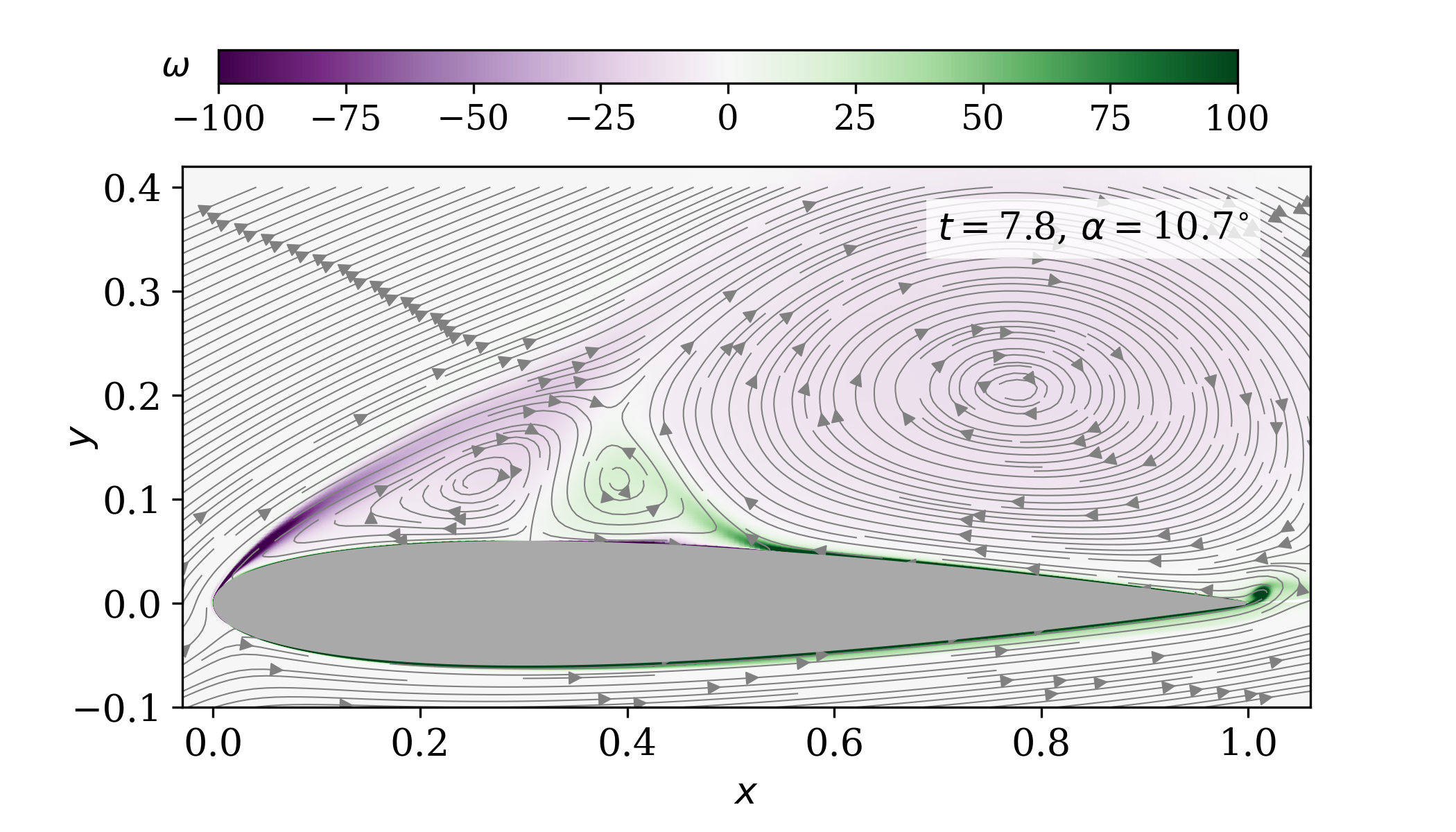}}
    \caption{Flow field vorticity contours overlaid with streamlines for the baseline case without control (a \& d), control using $\sigma_{\abs{BEF}}$ (b \& e), and  $\sigma_{LESP}$ (c \& f), respectively, for Case B, at two different times. The top row is at $t=6$ and the bottom row is at  $t=7.8$.
    \label{fig:flowfields_ctrl_caseB}}
\end{figure}

\Cref{fig:flowfields_ctrl_caseB} presents snapshots of vorticity contours overlaid with streamlines for the baseline case, control applied using $\sigma_{\abs{BEF}}$ and control applied using  $\sigma_{LESP}$.
The flow fields are shown at two time instances after the stall criterion is met, namely, at $t=6$ (top row) and  $t=7.8$ (bottom row).
The cases where control is applied still show a DSV being shed, but it is significantly weaker compared to the baseline case.

\begin{figure}[htpb]
    \incfig[width=0.7\textwidth,trim=0.2cm 0.5cm 0.5cm 0.2cm,clip=true]{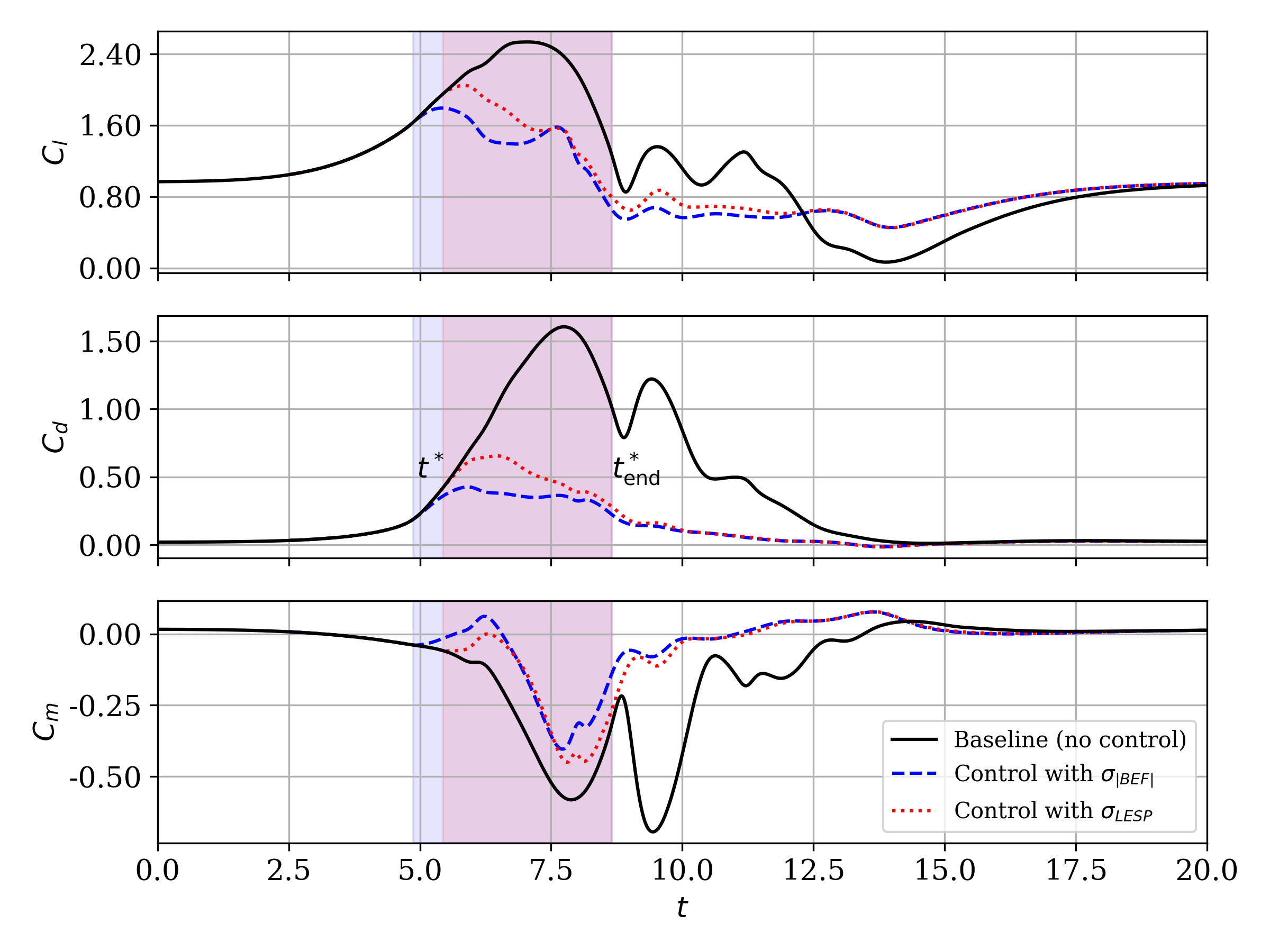}
    \caption{Time histories of the aerodynamic lift, drag, and moment coefficients -- $C_l$,  $C_d$, and $C_m$, for the uncontrolled (baseline) and control cases for Case B.}
    \label{fig:aerodyn_coeff_caseB}
\end{figure}

The top three panels of \cref{fig:aerodyn_coeff_caseB} show the time histories of the unsteady aerodynamic coefficients ($C_l$,  $C_d$, and $C_m$) for the baseline case and cases where control is applied.
The $\abs{BEF}$ and $LESP$ parameters suggest stall onset during the pitch-up phase of the motion ($t^* \sim 5$), with the $\sigma_{\abs{BEF}}$ criterion triggering earlier than $\sigma_{LESP}$.
The rapid increase in the aerodynamic coefficients (due to ensuing stall) is arrested and stall is mitigated via the proposed control strategy.
We do not claim that the control strategy presented here is optimal, but we argue that it successfully demonstrates the effectiveness of using stall indicators based on the $BEF$ and $LESP$ parameters to mitigate severe effects of dynamic stall.
Several enhancements to the control strategy are possible, including the use of closed-loop control using real-time sensor feedback and reinforcement learning, use of actuation, e.g., blowing/suction, plasma control, trailing edge flaps, etc. in place of controlling the maneuver, or a combination of these.

\section{Conclusion}
This work demonstrates the use of physics-based criteria utilizing the $LESP$ and $BEF$ parameters for detecting and mitigating unsteady stall in different types of airfoil motion. 
Two airfoil maneuvers, namely, a constant-rate pitch-up and a gaussian-modulated sinusoidal pitching, leading to stall, were simulated using the uRANS method.
A tracking parameter based on the variation of $\abs{BEF}$ and $LESP$ was used to trigger a control action.
For the constant-rate pitch-up maneuver, the control action involved gradually reducing $\alpha$ to a value below the static stall angle.
In the sinusoidal pitching case, $\alpha$ was first reduced to a value below static stall angle and gradually modified back to the original motion without risk of stall.
In both cases, the control action was triggered before the DSV was fully developed, using either the $BEF$ or $LESP$ parameters.
Both parameters were effective in mitigating the adverse effects (large variations in aerodynamic loads) of dynamic stall.
The $BEF$ parameter indicated stall onset earlier than the $LESP$ parameter, resulting in marginally improved control performance.
Future work could explore the integration of these physics-based criteria with data-driven methods, such as reinforcement learning, to develop more robust and efficient stall mitigation strategies that operate across a wide range of flight conditions. 

\bibliographystyle{plainnat}
\bibliography{main}

\begin{thebibliography}{13}
\providecommand{\natexlab}[1]{#1}
\providecommand{\url}[1]{\texttt{#1}}
\expandafter\ifx\csname urlstyle\endcsname\relax
  \providecommand{\doi}[1]{doi: #1}\else
  \providecommand{\doi}{doi: \begingroup \urlstyle{rm}\Url}\fi

\bibitem[Chandrasekhara(2007)]{Chandrasekhara2007}
M.~S. Chandrasekhara.
\newblock Compressible dynamic stall vorticity flux control using a dynamic camber airfoil.
\newblock \emph{Sadhana}, 32:\penalty0 93--102, 2007.
\newblock \doi{10.1007/s12046-007-0008-8}.
\newblock URL \url{https://doi.org/10.1007/s12046-007-0008-8}.

\bibitem[Economon et~al.(2016)Economon, Palacios, Copeland, Lukaczyk, and Alonso]{SU2}
Thomas~D. Economon, Francisco Palacios, Sean~R. Copeland, Trent~W. Lukaczyk, and Juan~J. Alonso.
\newblock Su2: An open-source suite for multiphysics simulation and design.
\newblock \emph{AIAA Journal}, 54\penalty0 (3):\penalty0 828--846, 2016.
\newblock \doi{10.2514/1.J053813}.
\newblock URL \url{https://doi.org/10.2514/1.J053813}.

\bibitem[Jameson(1991)]{jameson1991time}
A.~Jameson.
\newblock Time dependent calculations using multigrid, with applications to unsteady flows past airfoils and wings.
\newblock In \emph{10th computational fluid dynamics conference}, page 1596, 1991.

\bibitem[Liu et~al.(2025)Liu, Beckers, and Eldredge]{Liu2025}
Zhecheng Liu, Diederik Beckers, and Jeff~D. Eldredge.
\newblock Model-based reinforcement learning for control of strongly-disturbed unsteady aerodynamic flows, 2025.
\newblock URL \url{https://arxiv.org/abs/2408.14685}.

\bibitem[McCroskey(1981)]{McCroskey1981}
W.~J. McCroskey.
\newblock The phenomenon of dynamic stall.
\newblock Technical report, NASA, Washington, DC, 1981.

\bibitem[Menter et~al.(2003)Menter, Kuntz, and Langtry]{Menter2003}
F.~R. Menter, M.~Kuntz, and R.~Langtry.
\newblock \emph{{T}en {Y}ears of {I}ndustrial {E}xperience with the {SST} {T}urbulence {M}odel}, pages 625--632.
\newblock Begell House, Inc., 2003.

\bibitem[Narsipur et~al.(2020)Narsipur, Hosangadi, Gopalarathnam, and Edwards]{Narsipur2020}
Shreyas Narsipur, Pranav Hosangadi, Ashok Gopalarathnam, and Jack~R. Edwards.
\newblock Variation of leading-edge suction during stall for unsteady aerofoil motions.
\newblock \emph{Journal of Fluid Mechanics}, 900:\penalty0 A25, 2020.
\newblock \doi{10.1017/jfm.2020.467}.

\bibitem[O{\ss}wald et~al.(2016)O{\ss}wald, Siegmund, Birken, Hannemann, and Meister]{osswald2016l2roe}
Kai O{\ss}wald, Alexander Siegmund, Philipp Birken, Volker Hannemann, and Andreas Meister.
\newblock L2roe: a low dissipation version of roe's approximate riemann solver for low mach numbers.
\newblock \emph{International Journal for Numerical Methods in Fluids}, 81\penalty0 (2):\penalty0 71--86, 2016.

\bibitem[Ramesh et~al.(2014)Ramesh, Gopalarathnam, Granlund, Ol, and Edwards]{Ramesh2014}
K.~Ramesh, A.~Gopalarathnam, K.~Granlund, M.~V. Ol, and J.~R. Edwards.
\newblock Discrete-vortex method with novel shedding criterion for unsteady aerofoil flows with intermittent leading-edge vortex shedding.
\newblock \emph{Journal of Fluid Mechanics}, 751:\penalty0 500–538, 2014.
\newblock \doi{10.1017/jfm.2014.297}.

\bibitem[Sharma and Visbal(2019)]{Sharma2019}
A.~Sharma and M.~Visbal.
\newblock Numerical investigation of the effect of airfoil thickness on onset of dynamic stall.
\newblock \emph{Journal of Fluid Mechanics}, 870:\penalty0 870–900, 2019.
\newblock \doi{10.1017/jfm.2019.235}.

\bibitem[Sheng et~al.(2005)Sheng, Galbraith, and Coton]{Sheng2005}
W.~Sheng, R.~A.~McD. Galbraith, and F.~N. Coton.
\newblock {A New Stall-Onset Criterion for Low Speed Dynamic-Stall}.
\newblock \emph{Journal of Solar Energy Engineering}, 128\penalty0 (4):\penalty0 461--471, 11 2005.
\newblock ISSN 0199-6231.
\newblock \doi{10.1115/1.2346703}.
\newblock URL \url{https://doi.org/10.1115/1.2346703}.

\bibitem[Sudharsan et~al.(2022)Sudharsan, Ganapathysubramanian, and Sharma]{Sudharsan2022}
S.~Sudharsan, B.~Ganapathysubramanian, and A.~Sharma.
\newblock A vorticity-based criterion to characterise leading edge dynamic stall onset.
\newblock \emph{Journal of Fluid Mechanics}, 935:\penalty0 A10, 2022.
\newblock \doi{10.1017/jfm.2021.1149}.

\bibitem[Sudharsan et~al.(2023)Sudharsan, Narsipur, and Sharma]{Sudharsan2023}
S.~Sudharsan, S.~Narsipur, and A.~Sharma.
\newblock Evaluating dynamic stall-onset criteria for mixed and trailing-edge stall.
\newblock \emph{AIAA Journal}, 61\penalty0 (3):\penalty0 1181--1196, 2023.
\newblock \doi{10.2514/1.J062011}.
\newblock URL \url{https://doi.org/10.2514/1.J062011}.

\end{thebibliography}

\end{document}